# Arbitrary-order exceptional points from a single optomechanical cavity

Ning Wu[1,2], Kaiyu Cui[1,2,*], Ziming Chen[1,2], Chenxuan Wang[1,2], Xue Feng[1,2], Fang Liu[1,2], Wei Zhang[1,2,3], Hao Sun[1,2], Yongzhuo Li[1,2], and Yidong Huang[1,2,3,*]

[1] Department of Electronic Engineering, Tsinghua University, Beijing 100084, China

[2] Beijing National Research Center for Information Science and Technology (BNRist), Tsinghua University, Beijing 100084, China

[3] Beijing Academy of Quantum Information Sciences, Beijing, China

*Corresponding author: kaiyucui@tsinghua.edu.cn; yidonghuang@tsinghua.edu.cn

**Exceptional points (EPs)—non-Hermitian spectral singularities—shape the evolution dynamics ranging from topological mode control[1–4] to high-precision sensing[5–7]. Higher-order EPs unlock enriched and sharpened spectral topology beyond the minimal second-order counterpart, thus expanding capabilities in those domains. However, the intrinsic topological fragility of higher-order EPs has long hindered their robust experimental realization[8–11]. Here, we present a scalable architecture that implements arbitrary-order EPs via an optically mediated recurrent network comprising a single nanomechanical resonator and an ensemble of digital synthetic resonators. We experimentally validate this architecture's scalability for both linear and nonlinear mechanical EPs, demonstrating EPs up to the seventh order. Moreover, we reveal that the fundamental noise component, in contrast to some technical-noise contributions, shares the same system coupling channel and undergoes identical root-response amplification as the measured signal. The co-amplification mechanism sets the fundamental sensitivity limit near EPs of any order, as verified by direct signal-to-noise ratio measurements. Our work establishes a general platform for exploring higher-order EP-based phenomena while clarifying the fundamental boundary of non-Hermitian sensitivity enhancement. More broadly, it facilitates a connection between non-Hermitian physics and AI-inspired recurrent architectures across diverse physical systems.**

Non-Hermitian physics describes the dynamics of open systems that exchange energy with the environment, encompassing phenomena across multiple scales from particle decay[12], decoherence of photons and phonons[13], and collective dynamics of active matter[14] to black hole evolution[15]. In such non-Hermitian systems, exceptional points (EPs) emerge as nontrivial spectral singularities, where the eigenvalues and eigenvectors coalesce simultaneously[8–11,16]. Unlike Hermitian degeneracies, the EPs exhibit a Riemann-surface topology and an associated phase transition. These features not only result in peculiar dynamical behavior but also provide opportunities for developing diverse mode-control devices[1–4] and highly sensitive sensors[5–7]. Early EP research primarily focused on second-order EPs, the simplest form of exceptional degeneracy. Extending the research to higher-order EPs shows the potential to explore richer topological phenomena in high-dimensional systems[8] and enhance sensor performance near EPs[17]. For EP-based sensors, device performance is believed to be improved because the unique mode splitting scales with $\varepsilon^{1/N}$ near EPs. As the order $N$ of EPs increases, the system exhibits a sharper spectral response to the perturbation $\varepsilon$.

Motivated by these potential advantages, numerous theoretical models[18–21] have been proposed to realize higher-order EPs. In parallel, third-order EPs have been demonstrated in a variety of physical platforms, including optical[17], acoustical[22], optomechanical[23], spin[24], superconducting[25], and electrical[19,26] systems. These implementations typically rely on cascading multiple physical modes. As the system dimensionality increases, additional coupling pathways for fundamental noise (quantum and thermal noise) are introduced, and the control of system parameters becomes increasingly demanding. These obstacles, together with the fragile topology near higher-order EPs, constrain the scalability of such EPs, making their practical realization a significant challenge. More crucially, although the sharper spectral response near higher-order EPs is well established and has been experimentally confirmed[17], whether a sensor's fundamental sensitivity can be enhanced near EPs is currently under debate[6,7,26–32], mainly because the associated fundamental-noise dynamics are subtle. To date, the advantages of higher-order EPs for sensing applications remain elusive.

In this study, we propose a universal architecture for realizing arbitrary-order EPs through an optically mediated recurrent network consisting of a physical resonator and an ensemble of synthesized resonators implemented via digital computation. Using this architecture, we experimentally demonstrate nanomechanical EPs across multiple orders in both the linear and nonlinear (phonon-lasing) regimes by frequency topology mapping, including the realization of a seventh-order nonlinear EP (NEP). We further reveal that the fundamental noise component shares the same system coupling channel with the measured signal, resulting in the root-response co-amplification of both. This co-amplification mechanism sets the fundamental sensitivity limit near EPs regardless of their order. Consistently, this fundamental boundary is verified by signal-to-noise ratio (SNR) measurements near third- and seventh-order NEPs. At the same time, higher-order EPs can still provide sensing advantages in systems dominated by specific forms of technical noise, particularly when such noise is decoupled from EP dynamics. This work establishes a scalable and versatile platform for arbitrary-order EPs and reconciles the ongoing debate concerning fundamental sensitivity enhancement in non-Hermitian sensing. Our approach is transferable to diverse physical platforms, including spin[24], superconducting[25], and thermal[33] systems, paving the way for broader exploration of non-Hermitian topology and high-precision sensing with AI-based networks.

## Scalable architecture for arbitrary-order EPs

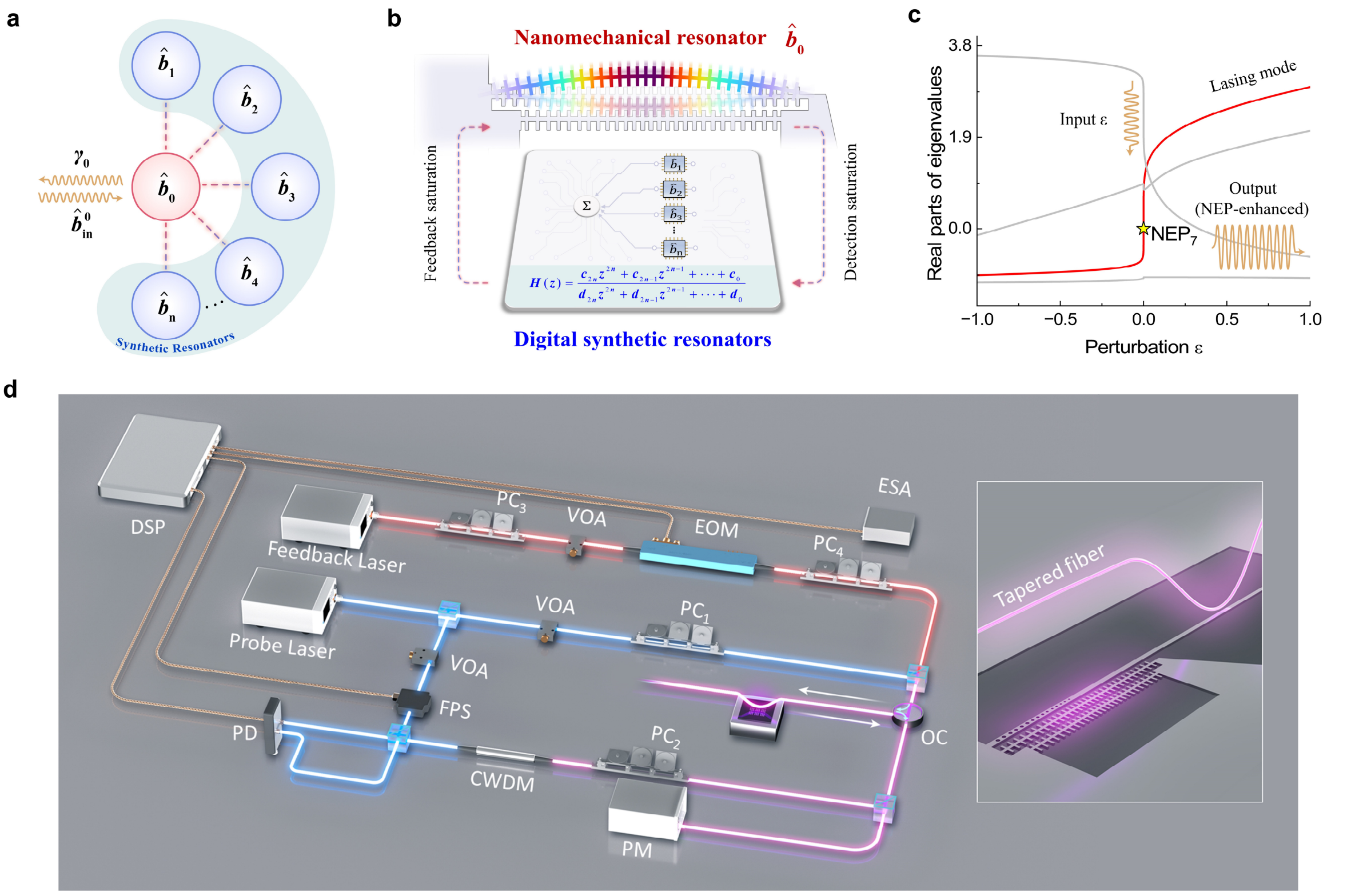

**Fig. 1. The proposed physical model and its realization platform.** (**a**) Physical model for realizing arbitrary-order EPs. $b_0$ is a real resonator that interacts with computationally defined synthetic resonators $b_1$ to $b_n$. The coupling strength between $b_0$ and $b_j$ is $J_j e^{i\varphi_j/2}$. The mode $b_0$ loses energy to the thermal bath at rate $\gamma_0$. (**b**) Physical platform for realizing arbitrary-order EPs. In this recurrent network, the nanomechanical resonator is coupled to an FPGA module, implementing the synthetic resonators. Feedback nonlinearity, arising from detection or output saturation, is used to realize NEPs. (**c**) Evolution of the real parts of eigenvalues near $NEP_7$ (star symbol) with a perturbation $\varepsilon$ applied to the frequency of the real resonator. Red and gray curves represent the lasing mode and other thermal modes, respectively. (**d**) Schematic of the experimental setup. PC: polarization controller; VOA: variable optical attenuator; EOM: electro-optic modulator; OC: optical circulator; PM: power meter; CWDM: coarse wavelength-division multiplexer; FPS: fiber phase shifter; PD: photodetector; ESA: electrical spectrum analyzer; DSP: digital signal processor.

The physical model for realizing arbitrary-order EPs is shown in Fig. 1a. The resonator mode $b_0$ is coupled to a set of modes $\{b_1,\cdots,b_n\}$. The complex coefficients $J_j e^{i\varphi_j/2}$ represent the coupling strength between $b_0$ and other modes $b_j$ $(j=1,\cdots,n)$. If the complex frequencies $\Omega_j - i\gamma_j/2$ ($j$=0…$n$) are set in advance, $J_j e^{i\varphi_j/2}$ can be precisely calculated by solving inhomogeneous linear equations to realize the EPs (Supplementary Information Section I). Conventionally, the direct way to construct such a higher-order EP is to couple multiple resonators in a real system. However, locating the singularity point within Riemann-surface topology becomes increasingly challenging as the dimensionality of the parameter space grows. Meanwhile, higher-order EPs are topologically fragile and exhibit extreme spectral sensitivity not only to the target signal but also to noise perturbations. Moreover, as the mode dimensionality increases, additional fundamental noise-coupling pathways are introduced into such systems, ultimately limiting the practical implementation and applications of higher-order EPs[9,34,35].

To overcome these obstacles, we implement a hybrid physical–digital architecture based on a recurrent network that supports EPs of arbitrary order, as shown in Fig. 1b. By leveraging the formal equivalence between computational processes and physical evolution, we reduce the physical realization to a single real resonator $b_0$ for sensing, while implementing all other resonators $b_j$ $(j=1,\cdots,n)$ and the interaction between $b_0$ and $b_j$ through digital computation. Experimentally, the silicon optomechanical fishbone cavity acts as the real resonator $b_0$, and an FPGA performs digital computation for feedback. The introduced digital computational process is immune to environmental perturbations, distinct from the analog implementation or the physical evolution process. Meanwhile, the thermal and quantum fluctuations, which scale with the number of physical resonators, are absent from the ideal computational process. Although the additional optical shot noise in the measurement process sets the displacement readout imprecision and induces radiation-pressure backaction on the mechanical mode[36,37], measurement noise at the order of mechanical zero-point motion is achievable[36–39]. In this study, the measurement noise and other technical contributions remain negligible compared with the thermal noise acting on the single real resonator $b_0$. Consequently, the challenges for realizing higher-order EPs are circumvented.

Furthermore, the system in Fig. 1b can enter the phonon lasing regime by increasing the coupling strength $J_j e^{i\varphi_j/2}$. In this regime, the feedback saturation persists until the mode with the highest gain reaches the gain-loss balance point[35,40]. As a result, the NEPs[26,28,30,41] can be realized by feedback saturation together with appropriately chosen system parameters. We propose a general method for realizing arbitrary-order NEPs through feedback saturation. Specifically, we predefine the system eigenvalues at NEPs of arbitrary order and solve inhomogeneous linear equations to obtain the corresponding complex frequencies $\Omega_j - i\gamma_j/2$ and coupling strengths $J_j^{NEP} e^{i\varphi_j/2}$ (Supplementary Information Section I). Accordingly, the FPGA coefficients are determined. Notably, the system eigenvalues at NEPs are not necessarily degenerate as in linear EPs (LEPs). This distinction arises because the system response of NEPs is not only determined by the instantaneous Hamiltonian[26] but also by the temporal dynamics of the system, which are controlled by the feedback saturation. When the perturbation $\varepsilon$ induces changes in the nanobeam's frequency or linewidth, the eigenfrequency shift depends not only on $\varepsilon$ but also on the relative feedback strength $\varsigma$. In the steady state, the coupling strength satisfies $J_j e^{i\varphi_j/2} = J_j^{NEP} e^{i\varphi_j/2}\sqrt{\varsigma}$, and $\varsigma$ is unity at NEPs. Moreover, because $\varsigma$ depends nonlinearly on $\varepsilon$, the order of NEPs can exceed the number of resonators. For $n$+1 resonators, NEPs up to the (2$n$)-*th* order, denoted as $\mathrm{NEP}_{2n}$, can be achieved. As the even-order NEPs are semi-stable for the perturbation $\varepsilon$ (Supplementary

Information Section I), we focus on the odd-order NEPs in this work. Fig. 1c displays the numerically calculated eigenvalue spectrum of a five-dimensional system, showing the non-degenerate eigenvalues at $NEP_7$. The lasing frequency shift in the vicinity of $NEP_7$ scales with $\sqrt[7]{\varepsilon}$ (Supplementary Information Section II).

## Characterization of higher-order EPs

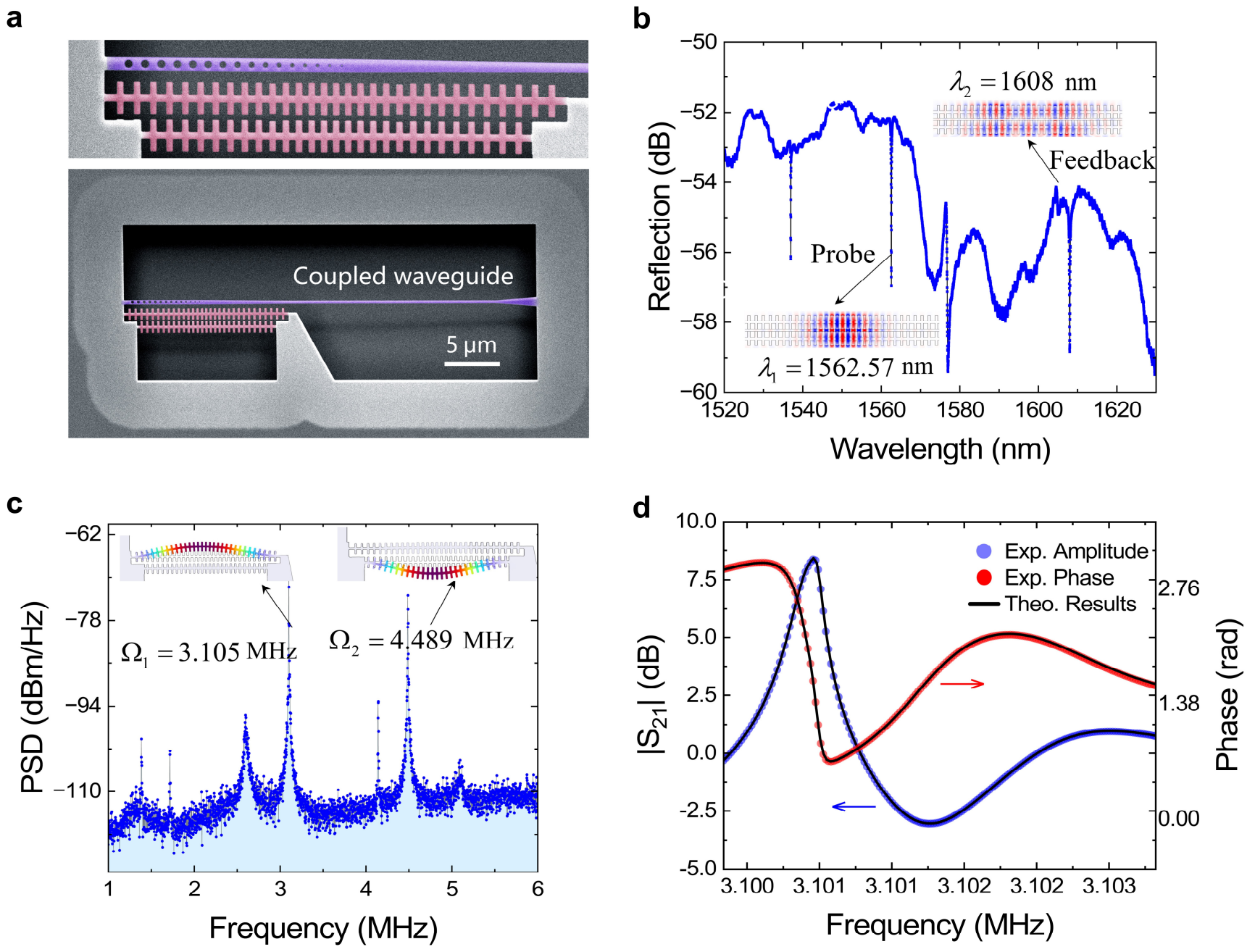

**Fig. 2. Experimental characterization of the optically mediated recurrent network.** (**a**) Scanning electron microscope (SEM) image of the fabricated silicon optomechanical cavity. The red and purple regions represent the fishbone cavity structure and the coupled waveguide, respectively. (**b**) Optical reflection spectrum. The cavity modes near 1563 nm and 1608 nm are used to detect the displacement of the nanomechanical resonator and to realize feedback of synthetic resonators, respectively. (**c**) Displacement power spectral density (PSD) of the mechanical motion in the MHz range. (**d**) $S_{21}$ curve of the FPGA-based synthetic-resonator module used to realize $NEP_7$.

To demonstrate the arbitrary-order LEPs and NEPs proposed above, we designed and fabricated a silicon optomechanical fishbone cavity (Supplementary Information Section III). The probe-feedback experimental setup in Fig. 1d was built to further demonstrate higher-order EPs (Supplementary Information Section IV). Fig. 2a shows the scanning electron microscope (SEM) image of the fabricated structure. The red area in Fig. 2a represents a fishbone nanobeam cavity that serves as the physical nanomechanical resonator for realizing EPs. Moreover, the periodic refractive index modulation near the boundary of the nanobeams localizes the optical field in the center of the defect region. The purple region is a reflective waveguide that couples light in and out through a tapered fiber attached above (Fig. 1d inset). The probe laser was used to excite one optical cavity mode to read out the displacement signal of the nanomechanical resonator via optomechanical interaction. Subsequently, the displacement signal was sent into the synthetic resonators implemented on the FPGA-based digital signal processor (DSP). The computational feedback of the synthetic-resonator module modulates the feedback laser that excites another optical mode, driving the mechanical mode via radiation pressure modulation.

Next, we present the basic characterization of the experimental system. The optical reflection spectrum is shown in Fig. 2b. The optical mode near 1563 nm with the fitted quality factor $Q_1 \approx 9977$ is driven by the probe laser, which is tuned to cavity resonance. The optical mode near 1608 nm with the fitted quality factor $Q_2 \approx 6030$ is driven by the feedback laser, whose wavelength is detuned by 1.5 nm from the cavity resonant wavelength, to minimize the amplitude-dependent saturation of the optical spring effect[13,35]. Meanwhile, the mechanical spectrum acquired from balanced homodyne detection[42] is shown in Fig. 2c. The peaks at 3.105 MHz and 4.489 MHz correspond to the fundamental modes of the cavity's two nanobeams. In this study, we focus on the mechanical mode with frequency $\Omega_0/2\pi \approx 3.105$ MHz and quality factor $Q_m \approx 2512$ for realizing higher-order EPs. Its optomechanical coupling rates to the probe and feedback optical modes are estimated as $g_1/2\pi = 0.912$ MHz and $g_2/2\pi = 1.1514$ MHz, respectively. The response of the DSP-constructed synthetic-resonator module is also characterized, as shown in Fig. 2d. The measured $S_{21}$ curve of this module realizing $NEP_7$ reproduces the theoretical response. The feedback path delay $t_0$ introduces a phase $\varphi_d$ in the coupling strength as $J_j e^{i(\varphi_j+\varphi_d)/2}$ $(j=1,\cdots n)$. Here, $\varphi_d \approx (\Omega_0 t_0 \mod 2\pi)$ and serves as an additional tuning parameter controlling the distance from the EPs ($\varphi_d = 0$).

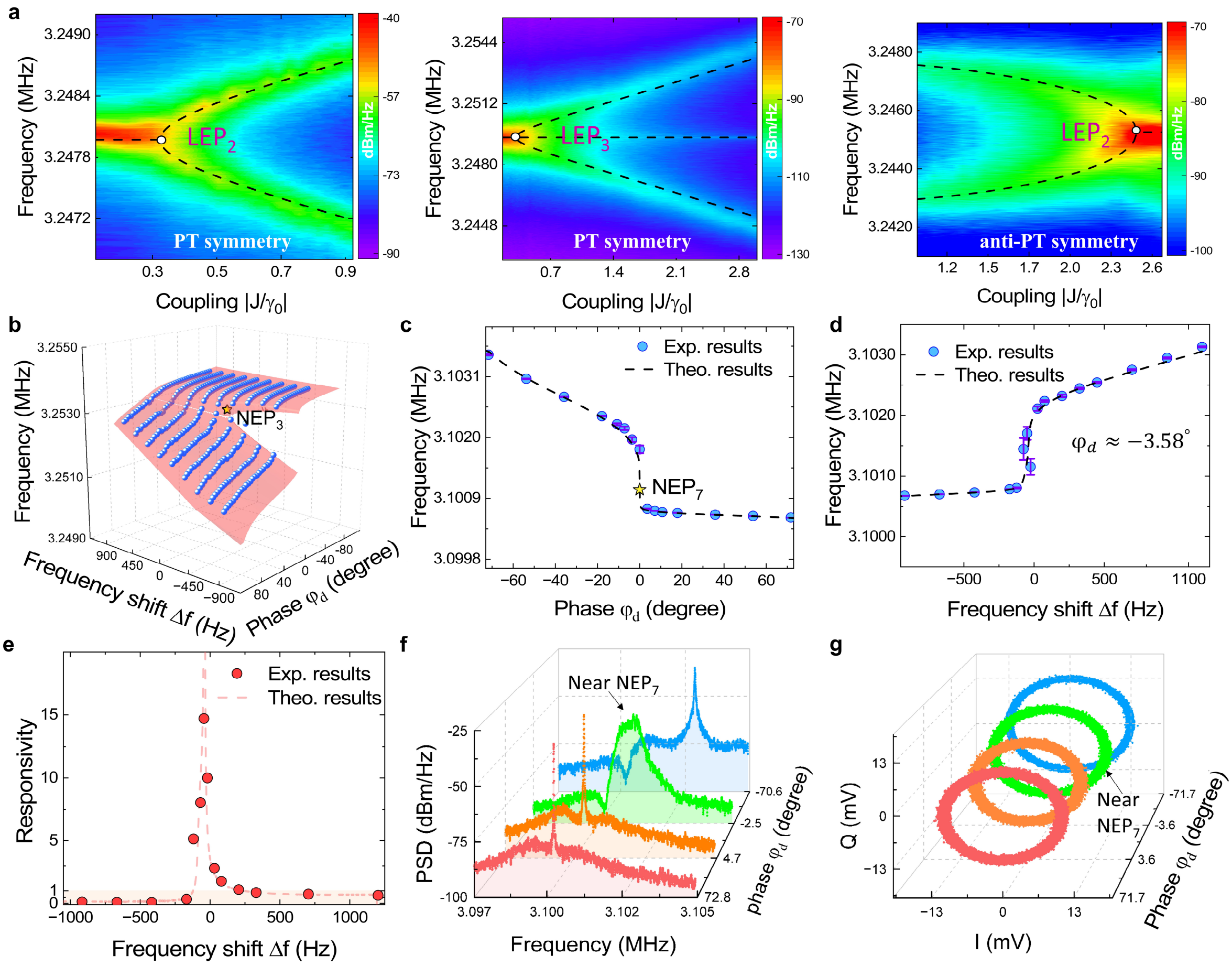

**Fig. 3. Experimental demonstration of higher-order EPs.** (**a**) Measured mechanical displacement PSD near LEPs, including PT-symmetric $LEP_2$/$LEP_3$ and anti-PT-symmetric $LEP_2$. (**b**) Lasing frequency evolution near $NEP_3$ as a function of the nanomechanical frequency shift $\Delta f$ and the delay phase $\varphi_d$. (**c**) Mechanical lasing frequency versus $\varphi_d$ near $NEP_7$ (star symbol). (**d**, **e**) Mechanical lasing frequencies (d) and the corresponding responsivities (e) versus $\Delta f$ at $\varphi_d \approx -3.58°$. In (e), the responsivities are derived by numerically differentiating

the experimental lasing frequencies, with the nanomechanical resonator replaced by a synthetic counterpart to improve accuracy. (**f**, **g**) Mechanical spectra near $NEP_7$ (f) and their complex amplitude representation (g) are shown at multiple $\varphi_d$. Dashed lines in (a) and surface in (b) represent theoretical results; colormap in (a) and dots in (b) show experimental data. (b, d) The $\Delta f$ sweep is equivalently implemented by uniformly shifting all synthetic-resonator frequencies $\Omega_j$. (c, d) The error bars denote the Allan deviation at $\tau$= 0.1 s and are doubled for clarity.

With this setup, higher-order EPs are demonstrated in Fig. 3. For simplicity, we begin by presenting the LEPs in Fig. 3a (Supplementary Information Section X). We construct the LEPs with both parity-time symmetry and anti-parity-time symmetry to show the flexibility of this proposal. In the parity-time symmetry configuration, the second- and third-order LEPs ($LEP_2$ and $LEP_3$) are illustrated by tuning the coherent coupling strength[35]. While the spectra near $LEP_2$ exhibit a square-root response, those near $LEP_3$ manifest a linear response as the system symmetry cancels the third-order response[17]. The second-order EP with anti-parity-time symmetry is also realized by tuning the dissipative coupling strength [35] between two mechanical modes with distinct frequencies.

To access the phonon lasing regime and demonstrate NEPs, we reconfigure the FPGA by increasing the feedback gain and resetting the coefficients of digital synthetic resonators, including the complex frequencies $\Omega_j - i\gamma_j / 2$ and the coupling strengths $J_j^{NEP} e^{i\varphi_j/2}$. Fig. 3b shows the evolution of phonon lasing frequencies near $NEP_3$ as functions of the delay phase $\varphi_d$ and the frequency shift $\Delta f$ of the nanomechanical resonator. In contrast to the LEP case, only a single phonon lasing mode emerges near NEPs. The experimental data fall on the theoretical surface, which originates from a branch of the Riemann surface.

To further demonstrate the scalability and generality of this method, the evolution of lasing frequencies near $NEP_7$ is characterized in Fig. 3c–d following the same procedure as for the $NEP_3$ case. In Fig. 3c, the phase is tuned to $\varphi_d \approx 0$ to realize the $NEP_7$. Subsequently, the evolution of the lasing frequencies with the nanomechanical frequency shift $\Delta f$ is shown in Fig. 3d. The phase $\varphi_d$ is adjusted to $-3.58^\circ$ to avoid the bistability region[43] for further noise analysis. This slight inconsistency between theoretical and experimental results in Fig. 3c–d arises from the drift of the nanomechanical resonant frequency[44] (Supplementary Information Section IX). Consequently, the responsivity $\partial\omega_{las}/\partial\varepsilon$, defined as the derivative of lasing frequency versus perturbation $\varepsilon$, cannot be determined accurately from Fig. 3d due to the frequency-drift-induced errors. To address this, we replace this real nanomechanical resonator with a digital one that is highly stable, and perform numerical differentiation of experimental lasing frequencies to approximate $\partial\omega_{las}/\partial\varepsilon$ (Supplementary Information Section IX). The resulting responsivity is shown in Fig. 3e and greatly exceeds unity near $NEP_7$. In contrast, Fig. 3f presents the mechanical spectra under the same conditions as in Fig. 3c and reveals a broader laser linewidth near $NEP_7$. As shown in Fig. 3g, the in-phase/quadrature (*I*-*Q*) mechanical trajectory acquired over 10 s forms a limit cycle near $NEP_7$, confirming that the nanobeam is in the phonon-lasing state and ruling out subthreshold amplification as the source of linewidth broadening. In fact, we attribute the observed linewidth broadening to the thermal and vacuum noise acting on the nanomechanical resonator. This interpretation is supported by the Allan deviation analysis in the following section and by the enhanced locking range observed near NEPs in the injection-locking experiment (Supplementary Information Section VI).

## Fundamental noise sets the sensing limit

Further quantitative analysis of the noise is presented in Fig. 4 to investigate the sensitivity near NEPs. Specifically, Fig. 4a shows the Allan deviation $\sigma(\tau)$[32] derived from the Fig. 3d datasets. The increased Allan deviation near $NEP_7$ corresponds to the linewidth broadening. Moreover, we find that the Allan deviation is measurement-port-dependent. The Allan deviation calculated from the measurement signal at the $b_0$ port, denoted as $\langle e_0 | \Psi \rangle$, deviates from the -1/2 slope expected at short averaging times $\tau$ on a log-log plot. If only the lasing mode is considered, the relative phase between the resonators will be constant, resulting in synchronization. Under this condition, the noise will be independent of the measurement port. Therefore, the observed port-dependent noise originates from the other modes below the lasing threshold. By filtering out other modes using the projection $\langle \Psi_L^{las} | \Psi \rangle$, where $\Psi_L^{las}$ is the estimated left eigenvector of the lasing mode, the Allan deviation follows the -1/2 slope for short $\tau$ (Supplementary Information Section V).

The corresponding PSD $S_{ff}$ of the measured instantaneous lasing frequencies further illustrates this noise behavior (Fig. 4b). For high-frequency components, the frequency-jitter PSDs remain nearly flat, corresponding to the -1/2 slope of the Allan deviation. For low-frequency jitter, the PSDs lie above their high-frequency counterparts, especially near $NEP_7$. On the surface, this behavior might suggest that the long-term frequency drift[44], when treated as a signal, is more strongly amplified at low frequencies than thermal noise, thereby implying a sensitivity enhancement near NEPs. A similar conclusion can be drawn by comparing the high-frequency noise floor in $S_{ff}$ with the responsivity (Supplementary Information Section VII). Nevertheless, we find that thermal noise can also lead to a frequency-dependent floor in $S_{ff}$, indicating that high-frequency jitter alone is insufficient to represent the fundamental noise. This unexpected finding raises a question of whether the fundamental sensitivity is truly enhanced near NEPs and how it scales with EP order. We note that the fundamental sensitivity is established not only by the fundamental noise, but also by the optimal protocol given a fixed resource (phonon number of mode $b_0$).

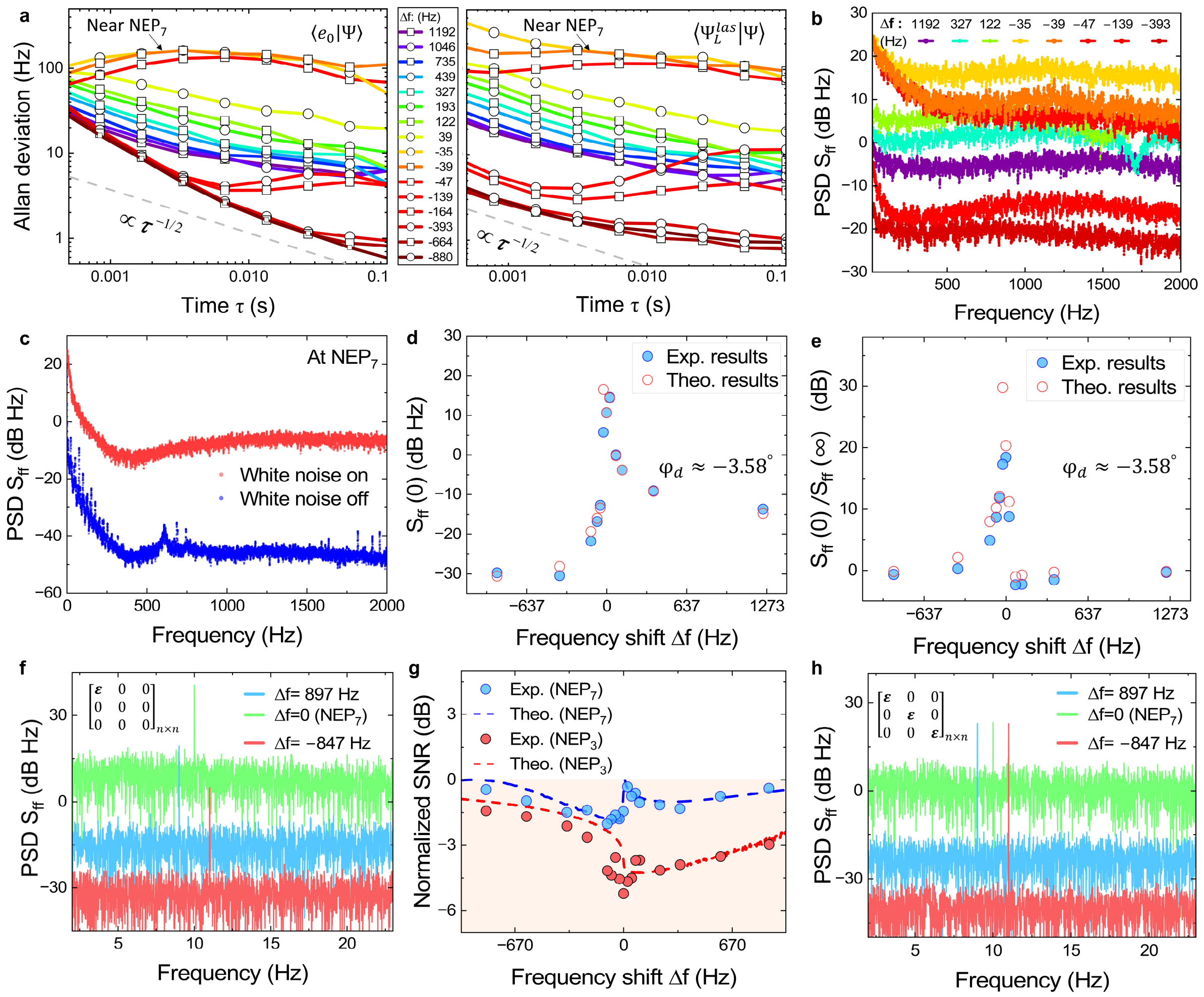

**Fig. 4. Sensitivity analysis of higher-order NEPs.** (**a**) Allan deviation under different projection bases. Gray dashed lines indicate a reference slope of -1/2. (**b**) PSD $S_{ff}$ of measured instantaneous mechanical lasing frequency near NEP7, under the projection $\left\langle \Psi_L^{las} \middle| \Psi \right\rangle$. (**c**) PSD $S_{ff}$ at NEP7 with and without additional white noise injected into port $b_{in}^0$. (**d, e**) Evolution of the PSD $S_{ff}$ versus the frequency shift $\Delta f$ of mode $b_0$ near NEP7 at $\varphi_d \approx -3.58°$, displaying $S_{ff}(0)$ at 0 Hz (d) and the relative value $S_{ff}(0)/S_{ff}(\infty)$ (e). (**f, h**) PSD near NEP7 at $\varphi_d = 0°$ after applying a 10 Hz perturbation $\varepsilon$ to the frequency of $b_0$ (f) and to all resonators (h). Spectra are horizontally offset for clarity. (**g**) Normalized SNR near higher-order NEPs at $\varphi_d = 0°$. The orange region indicates that the SNR is lower than that of a single-mode phonon laser with the same phonon energy in $b_0$. In (c-h), the nanomechanical resonator is replaced by a digital resonator for noise analysis, excluding the influence of environmental parameter drift.

We first verify the thermal-noise-induced frequency dependence of $S_{ff}$ in Fig. 4c, where we replace the nanomechanical resonator with a digital resonator to eliminate the frequency drift. Here, $S_{ff}$ increases dramatically when additional electrical white noise, serving as an equivalent thermal-noise source, is injected through the $b_0$ port. This indicates that white noise becomes the dominant noise source. Counterintuitively, the equivalent thermal noise leads to a frequency-dependent noise floor in $S_{ff}$ at NEP7. Specifically, the actual frequency jitter at low frequencies is significantly larger than that at high frequencies, which is not captured by the traditional Petermann factor[29]. This frequency-dependent jitter is further verified by Simulink simulations (Supplementary Information Section VIII) and can be explained by the frequency-dependent feedback response.

$$\frac{d\hat{b}_0}{dt} = (-i\Omega_0 - \gamma_0/2)\hat{b}_0 + \sum_{j=1}^{n} J_j^{NEP}\sqrt{\zeta}e^{i\varphi_j/2}\hat{b}_j + \sqrt{\gamma_0}\hat{b}_{in}^0 \tag{1}$$

To further reveal the fundamental sensitivity of the NEP-based sensors, we analyze the Langevin equation for the mechanical mode $b_0$, given by equation (1). Here, $J_j^{NEP}\sqrt{\zeta}e^{i\varphi_j/2}$ denotes the coupling strength between $b_0$ and the synthesized mode $b_j$, while $\sqrt{\gamma_0}b_{in}^0$ denotes the input noise drive from the thermal and vacuum fluctuations. For the high-frequency components of $\sqrt{\gamma_0}b_{in}^0$, the relative feedback strength $\zeta$ cannot track these components, as their frequencies exceed the response bandwidth. In contrast, $\zeta$ can adiabatically follow the low-frequency noise perturbations, making the measured $S_{ff}$ frequency-dependent. In the phonon lasing regime, the complex amplitude of mode $b_0$ takes the form $B(t)e^{-i\phi(t)}$. Therefore, the drive $\sqrt{\gamma_0}b_{in}^0$ is given by $\left(\sqrt{\gamma_0}b_{in}^0 e^{i\phi(t)} / B(t)\right)b_0(t) \approx \sqrt{\gamma_0 / 2}\left((X+iY)/\langle B(t)\rangle\right)b_0(t)$ . This noise drive is equivalent to the fluctuation of the mechanical resonant frequency as $\sqrt{\gamma_0 / 2}Y / \langle B(t)\rangle$ and mechanical dissipation rate as $\sqrt{2\gamma_0}X / \langle B(t)\rangle$, where $X$ and $Y$ are independent white-noise processes. As a result, the perturbation $\varepsilon$ to be measured, such as a shift in the mechanical frequency or linewidth, is mixed with a component of equivalent complex frequency jitter under an optimal protocol that distinguishes the contributions of the $X$ and $Y$ components. The signal and noise therefore share the same system coupling channel and are amplified simultaneously within the system's feedback response bandwidth, implying that the fundamental sensitivity is not enhanced near NEPs of any order. Overall, as the equivalence between the stochastic fundamental noise and dynamical parameter jitter, which underpins the signal-noise co-amplification, is general, the conclusion regarding the absence of fundamental sensitivity enhancement near EPs holds regardless of the EP order, the observable being measured, or whether the system is linear or nonlinear. For instance, the fundamental sensitivity near LEPs is discussed in Section V of the Supplementary Information, where we reach the same conclusion as in the NEP case.

The above conclusion is first verified experimentally by estimating the PSD $S_{ff}$ at 0 Hz in Fig. 4d and the relative PSD $S_{ff}(0) / S_{ff}(\infty)$ in Fig. 4e. These experimental results near $NEP_7$ are captured by the model that treats noise as equivalent parameter fluctuations (Supplementary Information Section V). We further measure the SNR to examine whether NEPs provide any fundamental sensitivity enhancement. A 10 Hz perturbation $\varepsilon$ is applied to the frequency of mode $b_0$ (Supplementary Information Section VIII), and the frequency shift $\Delta f$ of $b_0$ is swept across $NEP_7$ ($\varphi_d = 0$). As expected, the resulting PSD in Fig. 4f exhibits a peak at 10 Hz, well above the frequency-noise floor. However, no evidence of sensitivity enhancement near $NEP_7$ is observed. Furthermore, the measured SNRs near $NEP_3$ and $NEP_7$, normalized to the SNR of a single-mode laser with the same energy in $b_0$, are shown in Fig. 4g. These experimental results are in good agreement with the theoretical formula (Supplementary Information Section V). Contrary to the prevailing expectation, the normalized SNR is below unity, indicating that the sensing performance of NEP-based sensors does not surpass that of a simple single-mode laser. Since the frequency fluctuations $\sqrt{\gamma_0 / 2}Y / \langle B(t) \rangle$ alone would yield a normalized SNR of unity, the observed reduction in SNR stems from additional dissipation fluctuations $\sqrt{2\gamma_0}X / \langle B(t) \rangle$. The dissipation fluctuations generate redundant lasing frequency jitter, thereby making the enhancement of the signal slightly less than that of the noise near NEPs. These results support the conclusion that the higher-order NEPs do not provide fundamental sensitivity enhancement.

Although fundamental sensitivity enhancement is absent, the measured sensitivity can still be enhanced in the presence of additional technical noise[6,7]. Specifically, we apply a global frequency fluctuation equally to all resonators. This perturbation changes only the real part of the Hamiltonian's trace, without affecting the EP dynamics. As shown in Fig. 4h, this additional frequency fluctuation remains unamplified near $NEP_7$. This kind of technical perturbation is common in real systems. For instance, temperature and humidity fluctuations affect all resonators identically, and the probe laser frequency jitter in Ref. 7 effectively acts as a perturbation to the real part of the Hamiltonian's trace near the $LEP_2$. Consequently, higher-order EPs can still offer practical advantages for sensor applications in noisy environments through such non-fundamental sensitivity enhancement.

## Conclusion and outlook

In this study, we present a general architecture for realizing arbitrary-order EPs via a hybrid physical–digital recurrent network. With a single nanomechanical resonator, we experimentally demonstrate reconfigurable and scalable higher-order EPs in both linear and nonlinear regimes. Our approach addresses the intrinsic challenge posed by the topological fragility near higher-order EPs. It also obviates the need for both the resolved-sideband condition[23,35] and mechanical-parameter-mismatch compensation[45,46] when realizing EPs in optomechanical systems.

Furthermore, we investigate the sensitivity near EPs. Our analysis shows that the fundamental noise component and the signal enter the system through the same channel and are co-amplified. This mechanism indicates that the fundamental sensitivity is not enhanced near EPs regardless of their order, and this conclusion is supported by direct SNR measurements near $NEP_3$ and $NEP_7$. Despite the lack of fundamental sensitivity enhancement, the EPs remain effective for improving sensitivity in technical-noise-dominant systems[6,7]. Consequently, higher-order EPs can be practically useful for detecting physical quantities such as acceleration[6], spin[47], displacement[7], and mass[48].

In summary, our study demonstrates a general platform for realizing higher-order EPs and clarifies the long-standing debate on fundamental sensitivity enhancement in non-Hermitian sensors. Beyond its current realization, this programmable architecture opens avenues for exploring high-dimensional non-Hermitian topology and developing distributed sensing protocols that leverage exceptional points, quantum optomechanics, and artificial intelligence.

## Data availability

All data that support the findings of this study are available from the corresponding author upon reasonable request.

## Acknowledgements

The authors express their gratitude to Tianjin H-Chip Technology Group Corporation, Innovation Center of Advanced Optoelectronic Chip and Institute for Electronics and Information Technology in Tianjin, Tsinghua University for their fabrication support with electron beam lithography (EBL) and inductively coupled plasma (ICP) etching.

## Funding

This research was funded by the National Key R&D Program of China (Contract No. 2023YFB2806703); National Natural Science Foundation of China (Grant No. U22A6004); Beijing Frontier Science Center for Quantum Information; and Beijing Academy of Quantum Information Sciences.

## Author contributions

N.W. conceived the study and conducted the experiments. K.C. and Y.H. supervised the project. N.W. and K.C. analyzed the experimental results and wrote the paper. Z.C., C.W., X.F., F.L., W.Z., H.S., Y.L., and Y.H. discussed the results and reviewed the manuscript.

## Competing interests

The authors declare no competing interests.

Supplementary information for

# Arbitrary-order exceptional points from a single optomechanical cavity

Ning Wu[1,2], Kaiyu Cui[1,2,*], Ziming Chen[1,2], Chenxuan Wang[1,2], Xue Feng[1,2], Fang Liu[1,2], Wei Zhang[1,2,3], Hao Sun[1,2], Yongzhuo Li[1,2], and Yidong Huang[1,2,3,*]

*Corresponding author: kaiyucui@tsinghua.edu.cn; yidonghuang@tsinghua.edu.cn

## CONTENTS

## I. THEORETICAL MODEL OF ARBITRARY-ORDER EXCEPTIONAL POINTS

In this study, the arbitrary-order EPs are realized through a star-topology configuration, where the real resonator is coupled to multiple synthetic resonators. The effective Hamiltonian $H_{eff}$ of the system is expressed as:

$$\begin{pmatrix} \Omega_0 - i\gamma_0/2 & J_1 e^{i\varphi_1/2} & J_2 e^{i\varphi_2/2} & \cdots & J_n e^{i\varphi_n/2} \\ J_1 e^{i\varphi_1/2} & \Omega_1 - i\gamma_1/2 & 0 & 0 & 0 \\ J_2 e^{i\varphi_2/2} & 0 & \Omega_2 - i\gamma_2/2 & 0 & 0 \\ \vdots & 0 & 0 & \ddots & \vdots \\ J_n e^{i\varphi_n/2} & 0 & 0 & \cdots & \Omega_n - i\gamma_n/2 \end{pmatrix} \tag{S1}$$

where the diagonal elements $\Omega_j - i\gamma_j/2 \quad (j = 0,\cdots,n)$ denote the complex frequency of the resonator $j$. The off-diagonal elements $J_j e^{i\varphi_j/2}$ represent the coupling strength between the real resonator $b_0$ and the synthetic resonator $b_j$, and the phase $\varphi_j$ controls the ratio of dissipative to coherent coupling[1]. We shift the Hamiltonian to $H_{eff} - \omega_{ep} I$ without loss of generality, as this operation does not affect the dynamics. Here, $\omega_{ep}$ is defined as the eigenvalue at the linear EPs (LEPs) or the lasing frequency at the nonlinear EPs[2,3] (NEPs), depending on the operating regime. The eigenvalues $\Delta\omega$ of the new Hamiltonian are determined by the root equation:

$$\Delta\omega^{n+1} + p_n \Delta\omega^n + p_{n-1}\Delta\omega^{n-1} + \cdots + p_0 = 0 \tag{S2}$$

Here, the polynomial coefficient $p_n = -trace(H_{eff} - \omega_{ep} I)$, and $\left(p_0, p_1, \cdots, p_{n-1}\right)^T$ are affine functions of $\left(J_1^2 e^{i\varphi_1}, J_2^2 e^{i\varphi_2}, \cdots, J_n^2 e^{i\varphi_n}\right)^T$.

### A. Construction of LEPs

We first consider EPs in the linear regime. The polynomial coefficients from $p_0$ to $p_n$ must be set to zero to realize LEPs[4]. Here, the traceless Hamiltonian $H_{eff} - \omega_{ep} I$ implies $p_n = 0$. Since the conditions $\left(p_0, p_1, \cdots, p_{n-1}\right)^T = 0$ can be expressed as a set of inhomogeneous linear equations in $\left(J_1^2 e^{i\varphi_1}, J_2^2 e^{i\varphi_2}, \cdots, J_n^2 e^{i\varphi_n}\right)^T$, (n+1)-*th* order LEPs can be obtained by solving for the coupling strengths $J_j e^{i\varphi_j/2}$, provided that $\Omega_j - i\gamma_j/2$ are predetermined. If the perturbation $\varepsilon$ shifts the system away from the LEP condition, Eq. (S2) can be rewritten as

$$\Delta\omega^{n+1} + \frac{\partial p_n}{\partial \varepsilon}\varepsilon\Delta\omega^n + \frac{\partial p_{n-1}}{\partial \varepsilon}\varepsilon\Delta\omega^{n-1} + \cdots + \frac{\partial p_0}{\partial \varepsilon}\varepsilon = 0 \tag{S3}$$

The mode splitting can be further expressed as $\Delta\omega \approx \sqrt[n+1]{-\left(\partial p_0 / \partial \varepsilon\right)\varepsilon}$ if $\varepsilon$ is sufficiently small and $\partial p_0 / \partial \varepsilon \neq 0$.

## B. Construction of NEPs

When the system enters the phonon lasing regime where one or more eigenvalues have positive imaginary parts, feedback saturation occurs. This saturation will continue until the eigenvalue with the largest imaginary part achieves a gain-loss balance with zero imaginary component[1]. Therefore, when the perturbation $\varepsilon$ is applied to the real resonator $b_0$ at the NEPs, the coupling strength will be $J_j e^{i\varphi_j/2} = J_j^{NEP} e^{i\varphi_j/2}\sqrt{\varsigma}$ , where $\varsigma$ represents the relative feedback saturation strength. This implies that the shift of the lasing frequency depends not only on the perturbation $\varepsilon$ but also on the factor $\varsigma$. Accordingly, the real part of Eq. (S2) is given by:

$$\Delta\omega^{n+1} + \sum_{j=0}^{n} \Delta\omega^{j} \left( Re(p_j) + \frac{\partial Re(p_j)}{\partial \varepsilon}\varepsilon + \frac{\partial Re(p_j)}{\partial \varsigma}\delta\varsigma \right) = 0 \tag{S4}$$

In addition, the relationship between the factor $\delta\varsigma = \varsigma - 1$ and the perturbation $\varepsilon$ is determined by the imaginary part of Eq. (S2):

$$\delta\varsigma = -\left( \sum_{j=0}^{n} \Delta\omega^{j} \left( Im(p_j) + \frac{\partial Im(p_j)}{\partial \varepsilon}\varepsilon \right) \right) \left( \sum_{j=0}^{n} \Delta\omega^{j} \frac{\partial Im(p_j)}{\partial \varsigma} \right)^{-1} \tag{S5}$$

By substituting Eq. (S5) into Eq. (S4), the relation between the relative lasing frequency $\Delta\omega$ and the perturbation $\varepsilon$ can be expressed as:

$$\sum_{j=0}^{2n} C_j \Delta\omega^{j} = \sum_{j=0}^{2n} B_j \Delta\omega^{j}\varepsilon \approx B_0\varepsilon \tag{S6}$$

where $C_j$ and $B_j$ are defined in Eq. (S7).

$$\begin{aligned}
&C_j = A_j \;\; (0 \le j < n+1) \qquad C_j = A_j + \frac{\partial Im(p_{j-n-1})}{\partial \varsigma} \;\; (n+1 \le j \le 2n) \\
&A_j = \sum_{s+t=j} \left( Re(p_s)\frac{\partial Im(p_t)}{\partial \varsigma} - Im(p_s)\frac{\partial Re(p_t)}{\partial \varsigma} \right) \\
&B_j = \sum_{s+t=j} \left( -\frac{\partial Re(p_s)}{\partial \varepsilon}\frac{\partial Im(p_t)}{\partial \varsigma} + \frac{\partial Im(p_s)}{\partial \varepsilon}\frac{\partial Re(p_t)}{\partial \varsigma} \right)
\end{aligned} \tag{S7}$$

Therefore, the conditions for realizing $k^{th}$-order NEPs are:

$$C_j = 0 \text{ for } j < k \text{ and } C_k \neq 0 \tag{S8}$$

The coefficients $C_j$ are linear combinations of $\partial(p_j)/\partial\varsigma$, provided that the coefficients $p_j$ are predetermined. Here $C_0$=0, since $p_0 = 0$ when $\omega_{ep}$ is used as the reference frequency. Specifically, for the highest (2$n$)-*th* order NEPs with $n$+1 resonators, the complex parameters $\partial(p_j)/\partial\varsigma$ $(j = 0,\cdots,n-1)$ in a 2$n$-dimensional space are obtained by solving a 2$n$-by-2$n$ inhomogeneous system of linear equations, provided that the nonzero coefficient $C_{2n}$ is specified.

Practically, $\partial p_j/\partial\varsigma$ $(j = 0,\cdots,n-1)$ are controlled by the relative complex frequencies $\Omega_j - i\gamma_j/2 - \omega_{ep}$. $\partial p_j/\partial\varsigma = p_j - q_j$ and $q_j$ represents the corresponding polynomial coefficient of $\prod_{j=0}^{n}\left(\Delta\omega - (\Omega_j - i\gamma/2 - \omega_{ep})\right)$ for variable $\Delta\omega$. As the coefficients $p_j$ are predetermined, the complex resonator frequencies $(\Omega_j - i\gamma_j/2 - \omega_{ep})$ can be solved based on the relation between the coefficients $\partial p_j/\partial\varsigma$ and $q_j$. Then the vector of coupling coefficients $\left(J_1^2 e^{i\varphi_1}, J_2^2 e^{i\varphi_2}, \cdots, J_n^2 e^{i\varphi_n}\right)^T$ can be obtained using the methods for solving the LEPs, as the relative complex frequencies $(\Omega_j - i\gamma_j/2 - \omega_{ep})$ and the coefficients $p_j$ are all known.

The dimension of the constraints to realize the $k^{th}$-order NEPs in Eq. (S8) is $k$–1, which is smaller than the 2$n$-dimensional space of the complex coefficients $\partial(p_j)/\partial\varsigma$. Therefore, the $k^{th}$-order NEPs can be realized in a (2$n$–$k$+1)-dimensional subspace where the high-dimensional exceptional lines, surfaces, and volumes emerge naturally. In this study, we omit a detailed discussion of these high-dimensional structures and fix $C_j\,(2n-1 \le j \le 2n)$ to nonzero values to demonstrate (2$n$–1)-*th* order NEPs. More specifically, the complex frequency of the synthetic mode with an ultrahigh dissipation rate and its coupling to $b_0$ are predetermined. Thus, this mode can be adiabatically eliminated and treated as an additional complex frequency shift acting on the real resonator $b_0$. This means that the complex coefficient $\partial(p_{n-1})/\partial\varsigma$ is predetermined. As a result, the dimension of the equality constraints is equal to the dimension of $\partial(p_j)/\partial\varsigma$, which is 2$n$–2, sufficient to determine the unique system parameters $J_j e^{i\varphi_j/2}$ and $\Omega_j - i\gamma_j/2$.

### C. Even–odd stability rule

Near a $k^{th}$-order NEP, the relative lasing frequency is given by $\Delta\omega \approx \sqrt[k]{B_0\varepsilon / C_k}$ . For even-order NEPs, $\Delta\omega$ is real only when $B_0\varepsilon / C_k \geq 0$, since a negative number does not yield real roots when $k$ is even. For odd-order NEPs, the root equation always has a real solution regardless of the sign of $\varepsilon$. The lasing frequency must remain real if the system is at equilibrium. For even-order NEPs with $B_0\varepsilon / C_k < 0$, the lasing mode must transition to another branch, where the expression Eq. (S4) is no longer applicable. Thus, the lasing frequency evolves discontinuously near even-order NEPs but continuously near odd-order NEPs. For this reason, we mainly present the odd-order NEPs experimentally in the main text.

## II. SIMULATION OF EIGENVALUE SPECTRA NEAR EPs

### A. Eigenvalue simulation

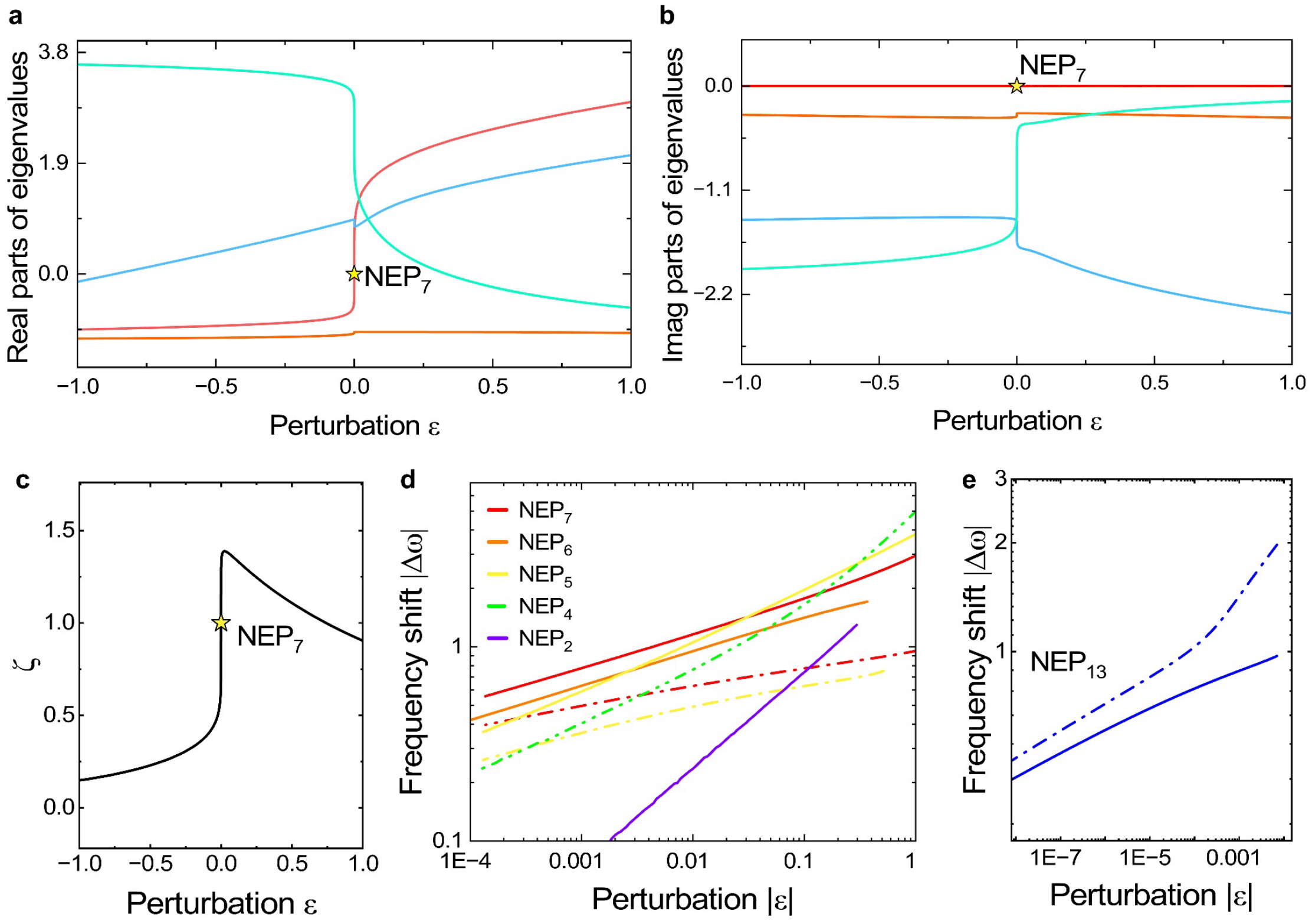

**Fig. S1. Numerical results for constructing NEPs using dimensionless parameters.** (**a** and **b**) The real and imaginary parts of eigenvalues near NEP$_7$ when a frequency perturbation $\varepsilon$ is applied to $b_0$. (**c**) The relative feedback strength $\varsigma$ as a function of perturbation $\varepsilon$. (**d**) The magnitude of frequency shift $|\Delta\omega|$ versus the perturbation strength $|\varepsilon|$ on a log-log scale for different-order NEPs constructed from a five-resonator system. (**e**) The frequency response on a log-log scale near $NEP_{13}$. The star symbol represents the complex phonon-lasing frequency in (a)-(b) and relative feedback strength in (c) at the $NEP_7$. In (d)-(e), solid and dash-dotted lines represent the frequency shift response under positive and negative perturbations, respectively.

As shown in Fig. S1, we simulate the eigenvalue evolution near NEPs based on the theoretical model in Section I. As an example, the $NEP_7$ is constructed from five resonators. Here, we adiabatically eliminate one resonator with the highest dissipation rate, and we specify four eigenvalues in the four-dimensional effective Hamiltonian. The corresponding eigenvalues at the $NEP_7$ are set to $-1-0.3i,\ 2.5-1i,\ 0.9-1.5i,$ and 0. The coupling between the eliminated mode and the real resonator $b_0$ shifts the complex frequency of $b_0$. For simplicity, we set the additional complex frequency shift to be $1i$.

Then the coefficients $\partial(p_j)/\partial\varsigma$ and the complex frequencies $(\Omega_j - i\gamma_j/2)$ are obtained by solving a six-by-six inhomogeneous system of linear equations. The complex frequencies are approximately $0.955-1.361i,\ 3.686-2.039i,\ -1.233-0.269i,$ and $-1.008-0.131i$, and the first value corresponds to the real resonator $b_0$. Correspondingly, the coupling strengths are solved from linear inhomogeneous equations similar to the method used for realizing LEPs in Section I, since both $p_j$ and $\Omega_j - i\gamma_j/2$ are known. Thus $J_1^{NEP}e^{i\varphi_1/2} \approx 0.937+1.637i,\ J_2^{NEP}e^{i\varphi_2/2} \approx 0.576+1.297i,$ and $J_3^{NEP}e^{i\varphi_3/2} \approx 0.0367-0.976i$. These parameters are used to realize $NEP_7$ in the main text. As for the $NEP_3$ in the main text, we set the eigenvalues of the effective Hamiltonian at the $NEP_3$ to $-1-0.3i$, and 0. The subsequent procedure is the same as that used for constructing $NEP_7$.

Fig. S1a–b shows the evolution of complex eigenvalues near the $NEP_7$ when the frequency perturbation $\varepsilon$ is applied to $b_0$. As expected, these eigenvalues do not coincide[2,3] with each other at $NEP_7$, and the phonon laser mode has zero imaginary part. Such a zero imaginary part is realized by feedback saturation. The relative feedback strength $\varsigma$ varies with the perturbation $\varepsilon$, as shown in Fig. S1c.

The core property of the higher-order NEPs is their characteristic root response. Fig. S1d shows the mode frequency shift magnitude under specific perturbation strength $|\varepsilon|$, plotted on a log-log scale. For the $NEP_7$, this simulated slope is 1/6.005 ($\varepsilon>0$) and 1/8.53 ($\varepsilon<0$) for the smallest perturbation, which closely matches the theoretical slope of 1/7. Moreover, other $k^{th}$-order NEPs can be realized by replacing the constraints $C_j = 0$ with other constants $C_j \neq 0$ where $k \le j < 7$ in Eq. (S7). The simulated results of other $k^{th}$-order NEPs in Fig. S1d also converge to the theoretical $1/k$ slope. Notably, for the even-order NEPs, the phonon laser frequency converges to zero only for one sign of the perturbation $\varepsilon$, unlike the odd-order cases. Furthermore, the $NEP_{13}$ is constructed using eight resonators, and the simulated slopes in the log-scale map are 1/13.1 for $\varepsilon<0$ and 1/14.7 for $\varepsilon>0$. The result shown in Fig. S1e verifies the validity of the method for constructing arbitrary-order NEPs.

## B. Digital realization of synthetic resonators

Since the complex frequency $\Omega_j - i\gamma_j/2$ and the coupling strength $J_j e^{i\varphi_j/2}$ are known, the feedback transfer function $H_F(\omega) = \sum_{j=1}^{n} -J_j^2 e^{i\varphi_j}\left((\Omega_j - \omega) - i\gamma_j/2\right)^{-1}$ is a rational function of $s$ with complex coefficients, where $s = -i\omega$. By contrast, the linear time-invariant transfer function described by the differential equations is a rational function of $s$ with real coefficients. This mismatch arises from the rotating-wave approximation, which is typically used in resonator modeling. Thus, to reconcile the complex transfer function with the standard form, $H_F(\omega)$ was approximated as

$$H_F(s) \approx -2\sum_{j=1}^{n} \frac{Im(J_j)^2 s^2/\omega_{ep} - 2sRe(J_j)Im(J_j) + Re(J_j)^2\omega_{ep}}{s^2 + \Omega_j^{\,2} + \gamma_j s} \tag{S9}$$

Here, the parameters are transformed as $J_j = J_j\gamma_0/\gamma_0'$, $\gamma_j = \gamma_j\gamma_0/\gamma_0'$ and $\omega_j = \omega_{ep} + \omega_j\gamma_0/\gamma_0'$ in advance. The parameters $\gamma_0$ and $\gamma_0'$ represent the dimensional and dimensionless mechanical dissipation rates, respectively. Here, the complex frequency of the mode that was adiabatically eliminated is set to $\omega_{ep} - i50\gamma_0$. As this linear time-invariant feedback is implemented digitally, we transfer the continuous function $H_F(s)$ to the discrete function $H_F(z)$ by using the impulse-invariant method. To avoid the coefficient sensitivity of high-order digital filters to numerical errors, $H_F(z)$ is expressed as a set of cascaded second-order sections:

$$H_F(z) = \prod_{j=1}^{n} \frac{c_{0j} + c_{1j}z^{-1} + c_{2j}z^{-2}}{d_{0j} + d_{1j}z^{-1} + d_{2j}z^{-2}} \tag{S10}$$

## C. System-level simulation of the higher-order EPs

Next, we discuss the detailed setup for realizing the EPs and simulating the behavior of the whole system in Simulink. The optomechanical coupling equations for the nanomechanical resonator are given by:

$$\begin{aligned} & m_{eff}\left(\frac{d^2x}{dt^2}+\gamma_0\frac{dx}{dt}+\Omega_0{}^2x\right)=F_{th}+F_{rp} \\ & \frac{da}{dt}=-\frac{\kappa}{2}a+i\Delta(x)a+\sqrt{\kappa_{ex}}\alpha_{in} \end{aligned} \tag{S11}$$

Here, the mechanical resonator with effective mass $m_{eff}$ is driven by the force $F_{th}$ from the thermal bath and the radiation-pressure force $F_{rp}$. Accordingly, the mechanical motion shifts the detuning $\Delta$ between the input laser and the optical cavity mode. As the optical dissipation rate $\kappa$ is much larger than the mechanical resonator frequency $\Omega_0$, the complex amplitude $a$ adiabatically follows the detuning $\Delta(x)$, yielding the steady-state amplitude $a\approx\sqrt{\kappa_{ex}}\alpha_{in}/\left(\kappa/2-i\Delta(x)\right)$. As a result, the output light field of the probe laser carries information about the mechanical motion, which is subsequently amplified via balanced homodyne detection (BHD). According to the input-output theory[5], the output electrical signal of BHD is proportional to $2\left|\alpha_A\right|(X_{BHD}\cos\theta+Y_{BHD}\sin\theta)$, where $\alpha_A$ denotes the amplitude of the local oscillator field injected from another port of the beam splitter in BHD. The $X_{BHD}$ and $Y_{BHD}$ quadratures are $X_{BHD}=\left|\alpha_{in}\right|\left(1-\left(\kappa_{ex}\times\kappa/2\right)/\left(\left(\kappa/2\right)^2+\Delta^2\right)\right)$ and $Y_{BHD}=-\left|\alpha_{in}\right|\kappa_{ex}\Delta/\left(\Delta^2+\left(\kappa/2\right)^2\right)$, respectively. The detuning $\Delta$ of the probe laser is set to zero to eliminate the optical spring effect[1]. Under this condition, the phase quadrature $Y_{BHD}$ reflects the mechanical motion, whereas the amplitude quadrature $X_{BHD}$ contributes negligibly. Thus, the relative phase $\theta$ between two laser beams is set to $\theta=\pi/2$ to filter $X_{BHD}$ quadrature and maximize sensitivity to the $Y_{BHD}$ quadrature. This analog electrical signal is digitized via an analog to digital converter (ADC) module and processed by a computational linear time-invariant system to implement interactions with other synthetic resonators. The resulting feedback signal $V$ then drives the electro-optic modulator (EOM) to generate intensity modulation of the feedback laser following $I_{EOM}\propto 1+\sin(\pi V/V_\pi)$, where $V_\pi$ is the half-wave voltage of the EOM and $I_{EOM}$ denotes the feedback-laser intensity. This modulated light intensity acts as the time-dependent radiation pressure force included in $F_{rp}$. Based on the above setup, the power spectral evolution as a function of the delay phase $\varphi_d$ near $NEP_3$ is obtained via Simulink, as shown in Fig. S2. The results follow the trajectory of one eigenvalue with zero imaginary part predicted by the theoretical model. Moreover, the linewidth broadening near NEP3 ($\varphi_d = 0$) is consistent with the experimental results presented in the main text.

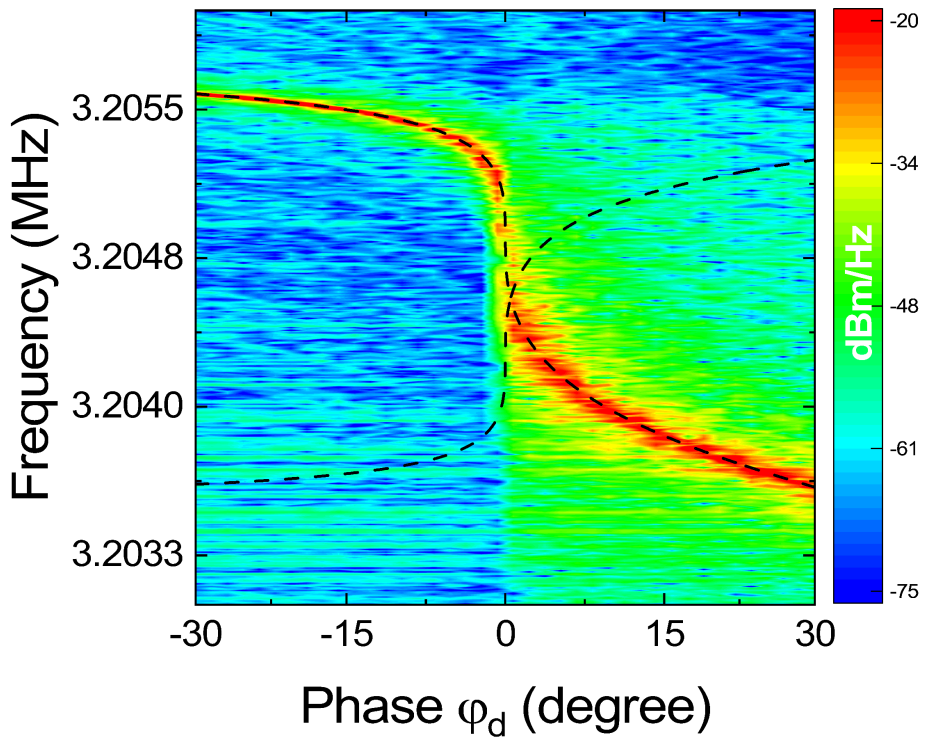

**Fig. S2.** Simulink simulation of the phonon laser spectral evolution versus the delay phase $\varphi_d$ near NEP3, compared with theoretical eigenvalues (dashed lines) under the rotating-wave approximation (RWA).

## D. Corrections for accurate realization of higher-order NEPs

The parameters used to implement NEPs in the main text are derived under the rotating-wave approximation, which is valid for our high-$Q$ resonator. Furthermore, we briefly discuss the general method for obtaining arbitrary-order NEPs beyond the RWA, showing the architecture's extensibility to low-$Q$ systems. The precise transfer functions of the real and the synthetic resonators are $H_0(s) = \left(s^2 + \Omega_0^{\,2} + \gamma_0 s\right)^{-1}$ and $H_F(s) = \varsigma \sum_{j=1}^{n} \left(s^2 + \Omega_j^{\,2} + \gamma_j s\right)^{-1} \left(q_{2j} s^2 + q_{0j} \Omega_j^{\,2} + q_{1j} \gamma_j s\right)$, respectively. The equilibrium of the system requires $H_0(s) H_F(s) = 1$, which can be further represented as:

$$(s + i\omega_{ep})^{2n+2} + r_{2n+1}(s + i\omega_{ep})^{2n+1} + \cdots + r_0 = 0 \tag{S12}$$

Here, Eq. (S12) is the general form of Eq. (S2), and the coefficients $r_j$ play the same role as $p_j$. Since the procedure for realizing arbitrary-order EPs beyond the RWA follows similar steps to those described in Section I, we omit the full procedure here. This method is further illustrated in Fig. S3. By using the parameters obtained from the RWA, the calculated practical lasing-frequency response (orange line) deviates from the ideal RWA model (blue line) at a representative finite mechanical quality factor. The red dots represent the results after correcting for the RWA-induced imprecision using Eq. (S12). The slope of these dots from the inset log–log plot converges to 1/3 for $NEP_3$, confirming the effectiveness of the method.

Moreover, the additional time delay $t_0$ is unavoidable in a real digital feedback system. Thus, the actual transfer function of the feedback system is $H_F(\omega)e^{-st_0}$. In the main text, we regard the influence of delay $t_0$ as the additional constant phase $\varphi_d$ because $st_0 \approx -i\omega_{ep}t_0$. When $\varphi_d = \omega_{ep}t_0 = 2m\pi$, this term is equivalent to zero-time delay. Nevertheless, the frequency-dependent correction term in the expansion $e^{-st_0} \approx e^{i\omega_{ep}t_0}(1+i(\omega-\omega_{ep})t_0)$ becomes non-negligible when $t_0$ is significant. The correction of the time delay required to accurately realize NEPs can be achieved similarly to the RWA correction above.

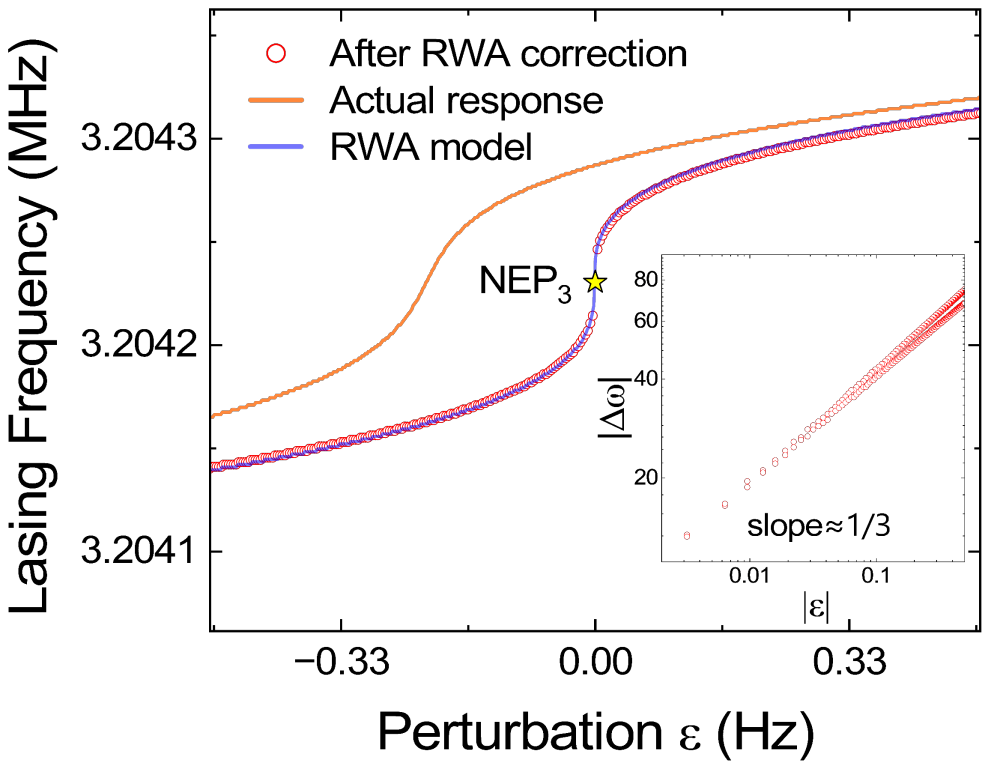

**Fig. S3.** Numerically calculated lasing-frequency response before (orange line) and after (red circles) applying the correction to the RWA-based parameters, compared with the theoretical RWA model (blue line). The yellow star marks $NEP_3$.

## III. DESIGN AND CHARACTERIZATION OF OPTOMECHANICAL CRYSTALS

We designed the fishbone optomechanical crystals in this study to demonstrate the higher-order EPs. In this design, the parameters of optomechanical crystals are shown in Fig. S4a. The period parameter P increases linearly from the center $P_1 = 380$ nm to the edge $P_{15} = 527$ nm. The parameter h is set to 200 nm, and it is crucial for determining the mechanical frequency. Parameters $L$=180 nm and $W_j$=1 μm ($j$=1–15) are initially chosen. $W_6$ is adjusted to 940 nm to introduce additional scattering loss to decrease the optical quality factor. The degraded optical quality factor helps reduce the influence of photothermal effects[6,7]. In principle, only one nanobeam is sufficient for demonstrating arbitrary-order EPs. The double beam structure with slot width S =160 nm is utilized to realize field enhancement in the slot so that the optomechanical coupling rate contributed from the moving-boundary effect[8] increases dramatically. This allows the mechanical motion to be measured with lower intracavity photon numbers, thereby further mitigating the photon-number-dependent detrimental effects. Moreover, the modes of each nanobeam are deliberately designed to have distinct frequencies to prevent their mutual interactions.

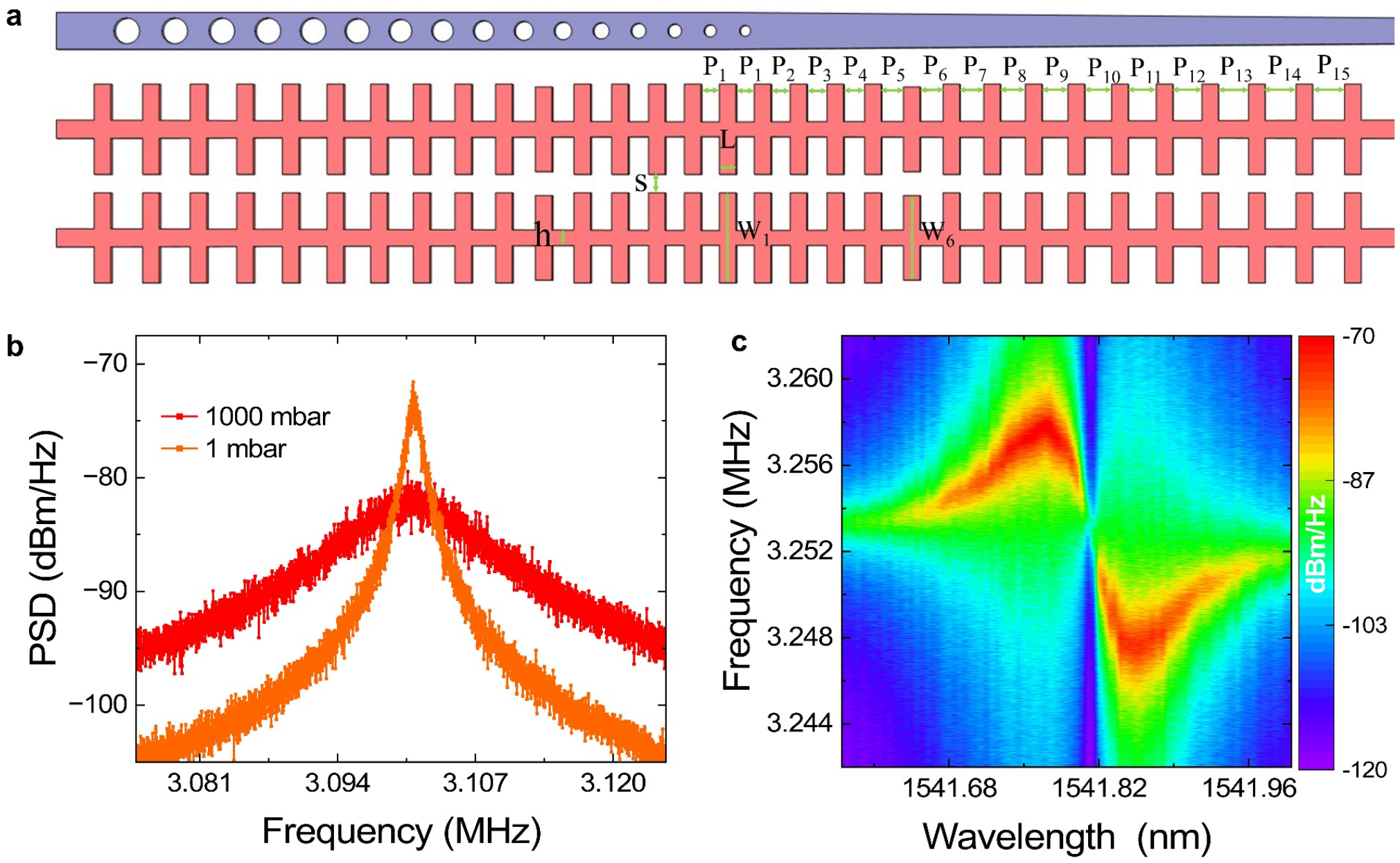

**Fig. S4. Design and preliminary characterization of fishbone optomechanical crystals.** (**a**) Parameters of the fishbone optomechanical crystals. (**b**) The mechanical displacement PSDs under different gas pressures. (**c**) The mechanical spectral shift versus the laser wavelength.

The structures are fabricated from a silicon-on-insulator wafer, which consists of a 220-nm-thick silicon top layer and a 3-μm-thick buried oxide layer. The fabricated sample is placed in the vacuum chamber to improve the mechanical quality factor, which is mainly limited by the gas damping[7] under ambient conditions. The corresponding spectra are presented in Fig. S4b, where the mechanical linewidth narrows with improved vacuum. The increase in the mechanical quality factor reduces the required feedback strength and intracavity photon number. Further decreasing the slot width S to enhance the optomechanical coupling rate[9] may enable robust realization of EPs under ambient conditions. After the sample is placed in the vacuum chamber at a pressure of ~$10^{-5}$ mbar, the optomechanical crystal is coupled with the pump light and the optical signal is directly detected by a photodetector (PD). Furthermore, the typical mechanical spectrum evolution by changing the detuning of the pump laser is shown in Fig. S4c. Clearly, such an optical spring effect is insignificant when the pump laser is resonant with the cavity near 1541.82 nm or far from resonance. Moreover, the experimental optomechanical coupling rates are deduced from the height of the thermal power spectral density (PSD)[1,10].

## IV. DETAILS OF EXPERIMENTAL SETUP

### A. DSP setup

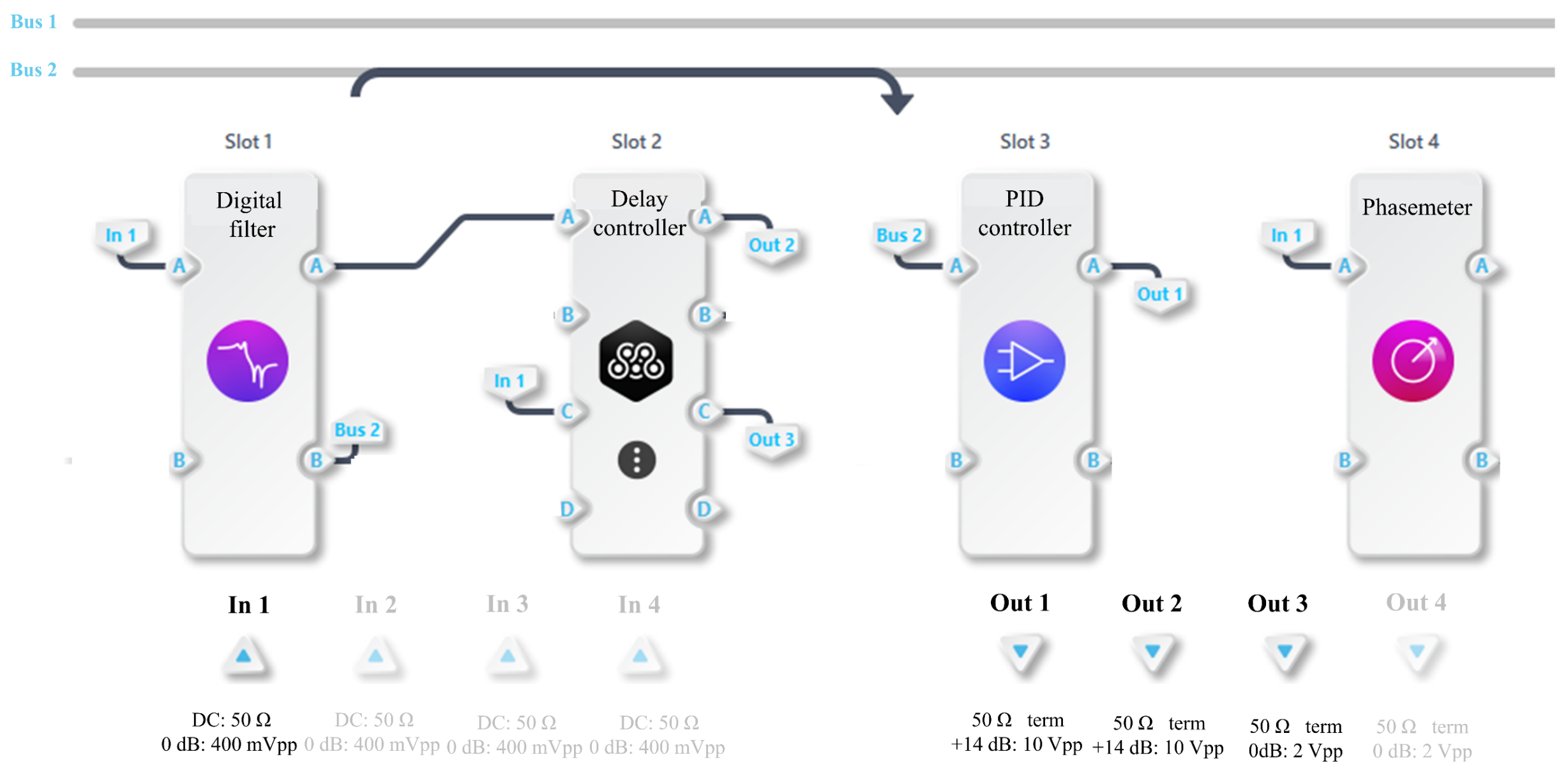

**Fig. S5.** Schematic of the DSP used to implement synthetic resonators and enable phase locking in balanced homodyne detection (BHD).

In our experiment, the probe light from the cavity was routed to a balanced homodyne detection setup[7,11,12], which converts the phase modulation into amplitude modulation and amplifies the mechanical signal further. The received light was converted into an electrical signal by a PD for further computational processing in an FPGA-based digital signal processor (DSP). Subsequently, the FPGA output, which was generated based on the multimode interaction, drove the EOM for feedback modulation. After that, the feedback laser was filtered using a coarse-wavelength-division multiplexer (CWDM) before the balanced homodyne detector. Moreover, the low-frequency signal from the PD was processed by the PID module[7,11,12] of the DSP, and then used to drive a fiber phase shifter (FPS) to stabilize the relative phase of the interference light against environmental fluctuations. Before data acquisition, the polarization controllers were tuned to minimize the polarization-dependent loss in the EOM ($pc_3$) or the optomechanical cavity ($pc_1$ and $pc_4$), and to maximize interference visibility ($pc_2$).

The digital signal processor (DSP) plays a crucial role in realizing higher-order EPs as it not only provides the feedback of synthetic resonators but also controls the BHD interference phase $\theta$, enabling high-precision measurement of mechanical motion. A schematic of the internal DSP architecture is shown in Fig. S5. The electrical signal from the photodetector (In 1) is processed by a digital filter to extract its low-frequency components and then sent to the PID controller, which generates the control signal (Out 1) to drive the phase shifter. In parallel, this same input signal is processed by another filter that generates the feedback of multiple synthetic resonators. The delay controller (shift register) and the cable length then adjust the signal delay to vary the phase $\varphi_d$, and this signal is amplified in the analog domain by 14 dB at the Out 2 port. Additionally, this signal is monitored by the phasemeter to record the frequency fluctuations of the phonon laser in the time domain. It is also sent to an electrical spectrum analyzer (ESA) at the Out 3 port to measure the mechanical displacement PSD.

## B. Feedback with saturation

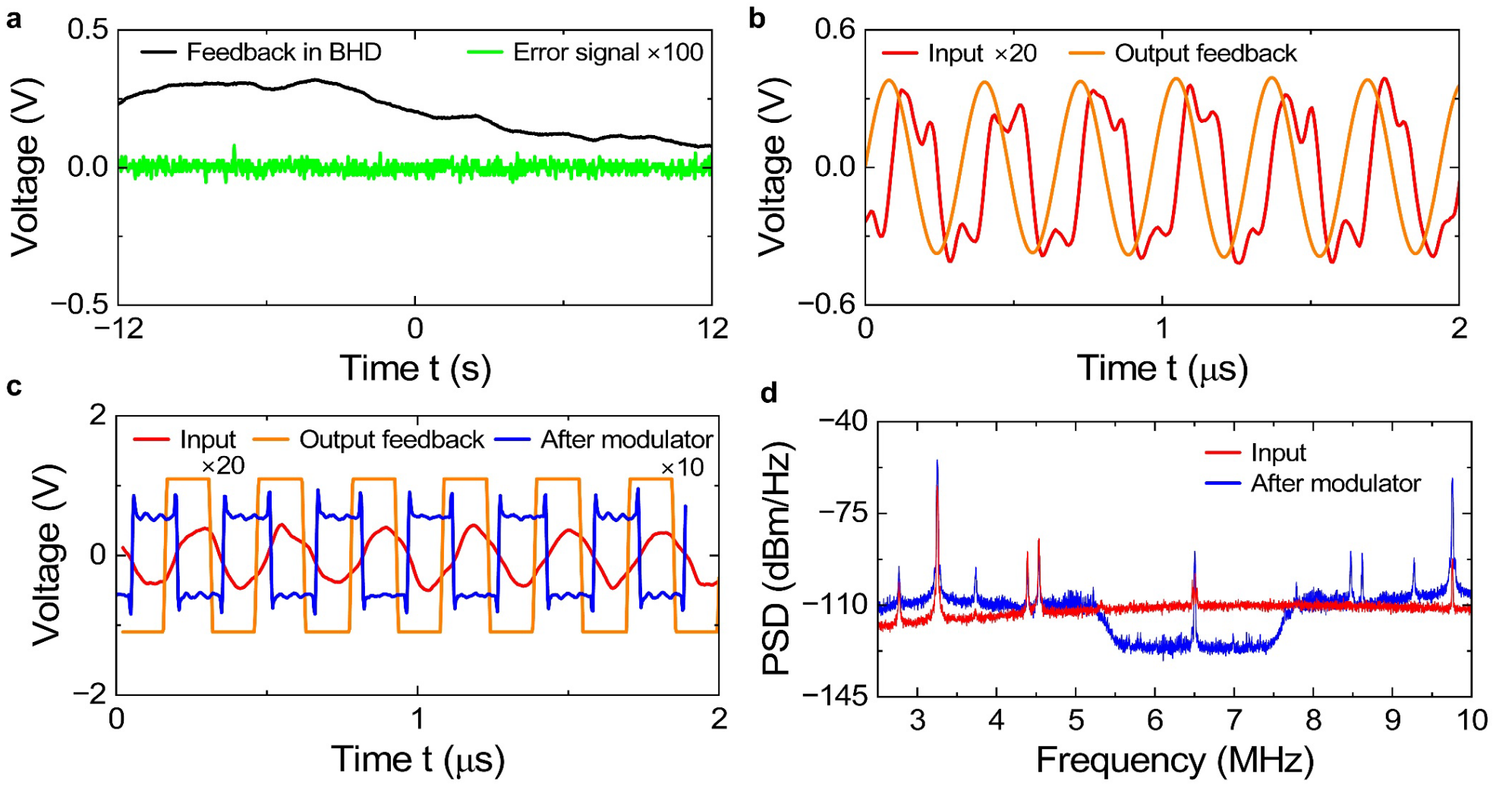

**Fig. S6. Experimental results of the feedback module.** (**a**) Feedback signal (black line) and error signal (green line) of the PID controller. (**b**) Input signal (red line) at the In 1 port and output signal (orange line) at the Out 2 port under measurement-induced saturation. (**c**) Input signal (red line) at the In 1 port and output signal (orange line) at the Out 2 port under digital limiter saturation. The blue line represents the radiation pressure signal after EOM modulation. (**d**) Frequency-domain representation of the blue and red signals shown in (c).

A typical feedback signal of the DSP is shown in Fig. S6. The feedback signal in the Out 1 port stabilizes the error signal at a value close to zero (Fig. S6a), thereby locking the interference phase $\theta$ to $\pi/2$. The feedback signal saturation, which is the core mechanism in the NEP experiments, is also determined by the DSP. For simplicity, this mechanism is demonstrated in a single-mode lasing experiment and shown in Fig. S6b–c. In our experiment, we identified three typical saturation mechanisms: measurement saturation, digital gain saturation, and modulation saturation. As shown in Fig. S6b, the measurement saturation results from the nonlinear response of the $Y_{BHD}$ quadrature to displacement $x$ as $Y_{BHD} = -|\alpha_{in}|\kappa_{ex}\Delta(x) / \left(\Delta(x)^2 + (\kappa/2)^2\right)$. In addition, this signal can also be digitally amplified after the interaction with synthetic resonators. Thus, output voltage limiting constitutes another saturation mechanism. The high voltage at the Out 2 channel can also cause modulation saturation of the EOM. The two additional saturation mechanisms in Fig. S6c can exist simultaneously with measurement saturation. It should be noted that the lasing frequency near NEPs is independent of these saturation mechanisms, although they can influence the response bandwidth to external perturbations.

In fact, the specific form of the saturation function $f(x)$ is insignificant for realizing NEPs as long as it is not an even function that would generate only even-order harmonics. For the odd nonlinear function of feedback saturation, its Taylor expansion is $f(x) = \sum_{j=0}^{\infty} \frac{f^{(2j+1)}(0)}{(2j+1)!} x^{2j+1}$. As the mechanical motion is $x = B(t)\cos(\omega_{las}t + \phi'(t))$, where $\phi'(t)$ denotes the relative phase drift, the term $x^{2j+1}$ can be expanded into odd harmonics $\cos\left((2k+1)\times(\omega_{las}t + \phi'(t))\right)$ with coefficients $\rho_{jk}$. Therefore, the output signal near the lasing frequency $\omega_{las}$ is $g(B)x(t)$, where $g$ is the nonlinear function of the amplitude $B$ as $g(B) = \sum_{j=0}^{\infty} \rho_{j0} \frac{f^{(2j+1)}(0)}{(2j+1)!} B^{2j}$. Moreover, it also reveals that only the odd-order harmonics of the fundamental frequency are allowed theoretically in the feedback output. The corresponding experimental result shown in Fig. S6d, where the saturation of the digital limiter is utilized, is consistent with the statement. Therefore, the DC component for this odd-function saturation remains zero under different lasing amplitudes $B$. It indicates that the pump laser power after the modulator in the feedback loop and the input DC signal for the PID controller are both irrelevant to the lasing amplitude.

## C. Measuring the average phonon number $n_L$

The PSD of the $Y_{BHD}$ quadrature is monitored to extract the phonon number $n_L$. In the measurement-saturation regime, where $Y_{BHD} = \kappa_{ex}\Delta(x)/\left(\Delta(x)^2 + (\kappa/2)^2\right)$, the third-to-first harmonic ratio follows a polynomial dependence on the amplitude $B$ as $\sum_{j=0}^{\infty} B^{2j} \frac{f^{(2j+1)}(0)}{(2j+1)!}\rho_{j1} / \sum_{j=0}^{\infty} B^{2j} \frac{f^{(2j+1)}(0)}{(2j+1)!}\rho_{j0}$. Therefore, the relative amplitude, $2\Delta(B)/\kappa$, can be extracted. As shown in Fig. S7, the experimental results of the normalized first- and third-harmonic intensities in a single-mode phonon laser are in good agreement with theoretical predictions. Meanwhile, the transduction coefficient relating $n_L$ to $2\Delta(B)/\kappa$ is calibrated using the thermal-motion reference of the mechanical mode, enabling extraction of $n_L$.

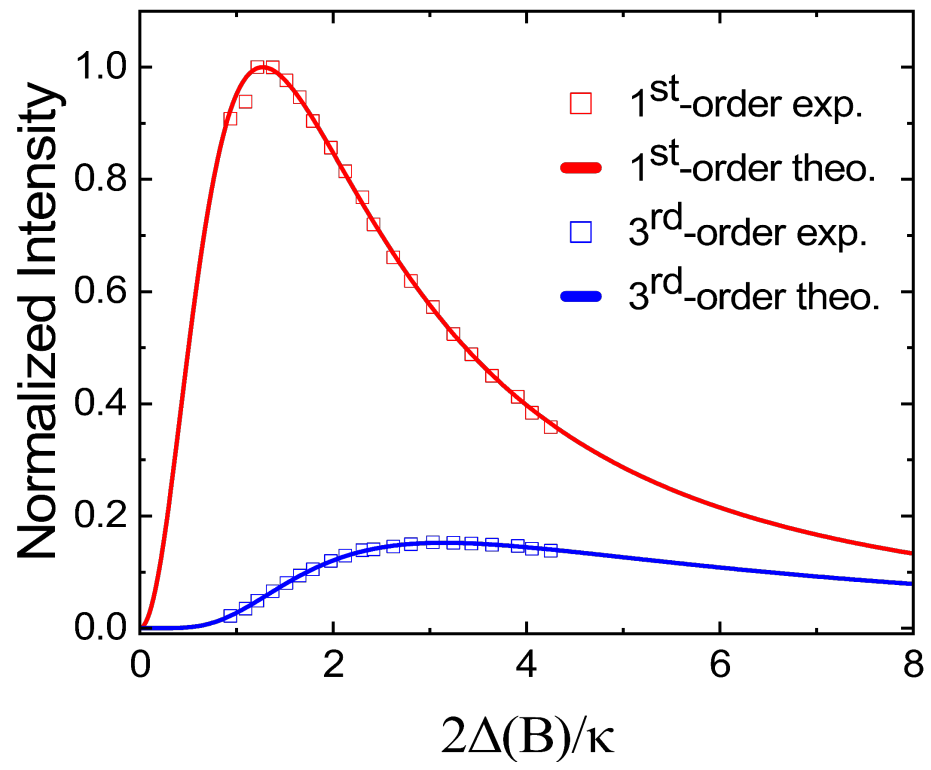

**Fig. S7.** Comparison between experimental and theoretical results for the normalized intensities of the first- and third-order harmonics in the PSD of the $Y_{BHD}$ quadrature.

During the measurement of the displacement PSD of the phonon laser, we observe that additional noise and instability arise when the lasing amplitude is sufficiently large and the optical mode is in the over-coupled regime. In the over-coupled regime, the DC component of $X_{BHD}$ quadrature $\left(1-(\kappa_{ex}\times\kappa/2)/\left((\kappa/2)^2+\Delta^2\right)\right)$, which serves as an error signal, will decrease or even reverse its sign as the mechanical amplitude increases. Therefore, PID locking instability causes performance degradation as described above. By contrast, the DC component of $X_{BHD}$ quadrature increases with the lasing amplitude in the under-coupled regime, and the PID locking remains robust against variations in the mechanical amplitude. Accordingly, the Allan deviation analysis in the main text is conducted when the optical mode for mechanical motion detection is in the under-coupled regime. Although the under-coupled and over-coupled optical modes can exhibit the same optical transmission spectrum[13], we can control the coupling between the waveguide and the cavity by rotating the tapered fiber attached to the waveguide. Here, a piezoelectric stage is used to vertically shift the SOI chip and rotate the tapered fiber. Consequently, the transition from the under-coupled regime to the over-coupled regime can be observed, accompanied by an increased optical linewidth. Meanwhile, the delay phase $\varphi_d$ shifts by approximately 161 degrees in realizing $NEP_3$. This phase shift further confirms the transition, as the corresponding locking phase $\theta$ is expected to reverse from $\pi/2$ to $-\pi/2$ theoretically.

## V. THEORETICAL SENSITIVITY ANALYSIS OF EPS

### A. Discussion of fundamental sensitivity near NEPs

The dynamics of an NEP-based phonon laser are governed by the effective Hamiltonian $H_{eff}$. Its time evolution can be expressed as:

$$i\frac{\partial|\Psi\rangle}{\partial t}=H_{eff}|\Psi\rangle+i|\Psi_{in}\rangle \tag{S13}$$

where

$$|\Psi\rangle=(b_0,b_1...,b_n)^T,\ \ |\Psi_{in}\rangle=(\sqrt{\gamma_0}b_{in}^0,\sqrt{\gamma_1}b_{in}^1...,\sqrt{\gamma_n}b_{in}^n)^T \tag{S14}$$

$|\Psi_{in}\rangle$ represents the stochastic drive from the thermal reservoir[1]. First, we neglect the time dependence of $H_{eff}$ on long time scales due to feedback response, and discuss the frequency jitter on short timescales. The left and right eigenvectors of $H_{eff}$ satisfy:

$$H_{eff}|\Psi_j^R\rangle=\omega_j|\Psi_j^R\rangle,\ \ \langle\Psi_j^L|H_{eff}=\omega_j\langle\Psi_j^L|$$

$$\langle\Psi_i^L|\Psi_j^R\rangle=0,\ i\neq j \tag{S15}$$

Here, the left eigenvectors $\langle\Psi_j^L|$ and the right eigenvectors $|\Psi_j^R\rangle$ have biorthogonal relations[14]. With $|\Psi\rangle=b_{las}|\Psi_{las}^R\rangle+\sum b_{other}|\Psi_{other}^R\rangle$, projecting $|\Psi\rangle$ onto $\langle\Psi_{las}^L|$ yields the evolution of the lasing-mode amplitude $b_{las}$, which satisfies

$$i\frac{\partial b_{las}}{\partial t}=\omega_{las}b_{las}+i\frac{\langle\Psi_{las}^L|\Psi_{in}\rangle}{\langle\Psi_{las}^L|\Psi_{las}^R\rangle} \tag{S16}$$

Consequently, the instantaneous frequency of $b_{las}$ ($b_{las}=|b_{las}|e^{-i\phi}$) is given by:

$$\frac{d\phi}{dt}=\omega_{las}-\frac{1}{|b_{las}|}\mathrm{Im}\left(\frac{\left\langle \Psi_{las}^{L}\middle|\Psi_{in}\right\rangle}{\left\langle \Psi_{las}^{L}\middle|\Psi_{las}{}^{R}\right\rangle}e^{i\phi}\right) \tag{S17}$$

Eq. (S17) describes the frequency jitter on a short-time scale. For an ideal computational feedback system, the only noise source is the input noise $b_{in}^{0}$ in $|\Psi_{in}\rangle$, which results from the thermal and quantum fluctuations in the nanomechanical resonator. We further show that this fundamental noise can be treated as parametric fluctuations of $b_0$ to better understand the long-time behavior. The time evolution of $b_0$ in Eq. (S13) is expressed as:

$$\frac{d\hat{b}_0}{dt}=-(i\Omega_0+\gamma_0/2)\hat{b}_0+\sum_{j=1}^{n}J_j^{NEP}\sqrt{\varsigma}e^{i\varphi_j/2}\hat{b}_j+\sqrt{\gamma_0}\hat{b}_{in}^{0} \tag{S18}$$

The thermal driving term $\sqrt{\gamma_0}b_{in}^{0}$ is equivalent to $\sqrt{\gamma_0}\left(b_{in}^{0}e^{i\phi}/|b_0|\right)b_0$. By defining $b_{in}^{0}e^{i\phi}$ as $(X+iY)/\sqrt{2}$, the driving term $\sqrt{\gamma_0}b_{in}^{0}$ is equivalent to additional $\left(\sqrt{\gamma_0/2}\right)Y/|b_0|$ frequency fluctuations and $\left(\sqrt{2\gamma_0}\right)X/|b_0|$ dissipation rate fluctuations. In addition, their correlation functions satisfy[15]:

$$\begin{aligned}
&\left\langle \hat{X}(t)\hat{X}(t+\tau)\right\rangle=\left(n_{th}+1/2\right)\delta(\tau)\\
&\left\langle \hat{Y}(t)\hat{Y}(t+\tau)\right\rangle=\left(n_{th}+1/2\right)\delta(\tau)\\
&\left\langle \hat{X}(t)\hat{Y}(t+\tau)+\hat{Y}(t)\hat{X}(t+\tau)\right\rangle=0
\end{aligned} \tag{S19}$$

where < > denotes ensemble averaging and $n_{th}$ is the thermal-equilibrium phonon occupancy. By introducing these parametric fluctuations into the effective Hamiltonian $H_{eff}$, Eq. (S16) is further revised as:

$$i\frac{\partial b_{las}}{\partial t}+i\left(b_{las}\left\langle \psi_{las}^{L}\middle|\partial_t\middle|\psi_{las}^{R}\right\rangle+\sum b_{others}\left\langle \psi_{las}^{L}\middle|\partial_t\middle|\psi_{other}^{R}\right\rangle\right)=\omega_{las}(t)b_{las} \tag{S20}$$

Here, $|\psi_{las}\rangle$ denotes the time-dependent eigenvector and $|\Psi\rangle=b_{las}(t)\left|\psi_{las}^{R}(t)\right\rangle+\sum b_{other}(t)\left|\psi_{other}^{R}(t)\right\rangle$. The solution of $b_{las}(t)$ is expressed as:

$$b_{las}(t)=b_{las}(0)e^{\int_0^t -i\left(\omega_{las}(t)-i\left\langle \psi_{las}^{L}\middle|\partial_t\middle|\psi_{las}^{R}\right\rangle-i\sum\frac{b_{others}}{b_{las}}\left\langle \psi_{las}^{L}\middle|\partial_t\middle|\psi_{other}^{R}\right\rangle\right)dt} \tag{S21}$$

Accordingly, the instantaneous frequency is given by:

$$\frac{d\phi}{dt}=Re\left(\omega_{las}(t)-i\left\langle \psi_{las}^{L}\middle|\partial_t\middle|\psi_{las}^{R}\right\rangle-i\sum\frac{b_{others}}{b_{las}}\left\langle \psi_{las}^{L}\middle|\partial_t\middle|\psi_{other}^{R}\right\rangle\right) \tag{S22}$$

In the low-frequency limit, the contributions of the last two terms can be neglected as $\partial_t \left|\psi_{las}^{R}\right\rangle$ and $\partial_t \left|\psi_{other}^{R}\right\rangle$ tend to zero. As a result, the measured frequency reduces to $\omega_{las}(t)$. Here, the feedback strength adiabatically follows the complex-frequency fluctuations in $\Omega_0 - i\gamma_0/2$, leading to a zero imaginary part of $\omega_{las}(t)$. Consequently, the slow fluctuation in $\omega_{las}(t)$ is governed by Eqs. (S4)–(S5), where the thermal driving effectively acts as a complex frequency perturbation in the nanomechanical resonator.

In this study, we use three kinds of quantities to characterize the frequency jitter. The first one is the PSD $S_{ff}$ of the measured instantaneous frequency $f_{las}(t)$ (Hz). For the high-frequency fluctuations that the system cannot follow, the correlation function of $f_{las}(t)$ is determined by Eq. (S17) and

$$\left\langle f_{las}(t) f_{las}(t+\tau)\right\rangle = \frac{\left|\left\langle \Psi_{las}^{L} \middle| e_0\right\rangle\right|^2 \left|\left\langle \Psi_{las}^{R} \middle| e_0\right\rangle\right|^2 \gamma_0}{(2\pi)^2 \left|\left\langle \Psi_{las}^{L} \middle| \Psi_{las}^{R}\right\rangle\right|^2 4n_L} \left\langle \left\{ \left(\hat{b}_{in}^{0}\right)^{\dagger}(t), \hat{b}_{in}^{0}(t+\tau) \right\} \right\rangle \tag{S23}$$

Here, $e_0$ is the unit vector in the direction of this real mode $b_0$ and {} denotes the anticommutator. As $\left\langle \left\{ \left(\hat{b}_{in}^{0}\right)^{\dagger}(t), \hat{b}_{in}^{0}(t+\tau) \right\} \right\rangle = 2(n_{th}+1/2)\delta(\tau)$ and following the Wiener-Khinchin theorem, the double-sided PSD is obtained as $S_{ff} = \frac{\gamma_0 \left(n_{th}+1/2\right) P_{eff}}{2n_L (2\pi)^2}$, where

$$P_{eff} = \frac{\left|\left\langle \Psi_{las}^{L} \middle| e_0\right\rangle\right|^2 \left|\left\langle \Psi_{las}^{R} \middle| e_0\right\rangle\right|^2}{\left|\left\langle \Psi_{las}^{L} \middle| \Psi_{las}^{R}\right\rangle\right|^2} \tag{S24}$$

For the low-frequency part of thermal driving, the lasing frequency $f_{las}(t) = \omega_{las}(t)/2\pi$ responds to the equivalent complex frequency fluctuations in $\Omega_0 - i\gamma_0/2$. The corresponding responsivities $\partial\omega_{las}/\partial\Omega_0$ and $\partial\omega_{las}/\partial\gamma_0$ are controlled by Eqs. (S4)–(S5). Therefore, combining Eq. (S19), the double-sided PSD $S_{ff}$ is expressed as:

$$S_{ff}(f \to 0) = \left( \left( \frac{\partial \omega_{las}}{\partial \Omega_0} \right)^2 + \left( \frac{\partial \omega_{las}}{\partial \left(\gamma_0/2\right)} \right)^2 \right) \frac{\gamma_0 (n_{th}+1/2)}{2n_L (2\pi)^2} \tag{S25}$$

An equivalent time-domain representation for analyzing the frequency jitter is the Allan deviation $\sigma(\tau)$ [14], complementary to the PSD $S_{ff}$. It is defined as:

$$\sigma(\tau) = \sqrt{\frac{1}{2} \left\langle \left[ \bar{f}(t+\tau;\tau) - \bar{f}(t;\tau) \right]^2 \right\rangle}, \quad \bar{f}(t;\tau) = \frac{1}{\tau} \int_t^{t+\tau} f(t') dt'. \tag{S26}$$

The Allan deviation $\sigma(\tau)$ characterizes the local spectrum of $S_{ff}$ [16], and the relationship between $\sigma(\tau)$ and the double-sided PSD $S_{ff}$ is:

$$\sigma(\tau)^2 = 2\int_{-\infty}^{+\infty} S_{ff}(f)\frac{\sin^4(\pi f \tau)}{(\pi f \tau)^2}df \tag{S27}$$

Specifically, for the white-noise frequency jitter, $\bar{f}(t;\tau) \sim N(0, S_{ff}/\tau)$ according to the central limit theorem. Consequently, $\sigma(\tau) = \sqrt{S_{ff}/\tau}$ .

The white-noise frequency jitter can alternatively be characterized by the phonon lasing linewidth. The phase noise $\phi$ leads to linewidth broadening, and the phase evolution in Eq. (S17) is described by a random walk process. The relative phase drift $\phi' = \phi - \omega_{las}t$ follows a normal distribution, with variance given by:

$$\left\langle \left(\phi'(t)-\phi'(0)\right)^2 \right\rangle = \frac{1}{\left|b_{las}\right|^2}\left\langle \int_0^t \mathrm{Im}\left(\frac{\left\langle \Psi_{las}^L \middle| \Psi_{in}(t)\right\rangle e^{i\phi'(t)}}{\left\langle \Psi_{las}^L \middle| \Psi_{las}^R\right\rangle}\right)dt \int_0^t \mathrm{Im}\left(\frac{\left\langle \Psi_{las}^L \middle| \Psi_{in}(t')\right\rangle e^{i\phi'(t')}}{\left\langle \Psi_{las}^L \middle| \Psi_{las}^R\right\rangle}\right)dt' \right\rangle = \frac{\gamma_0(n_{th}+1/2)P_{eff}}{2n_L}|t| \tag{S28}$$

Therefore, the PSD $S_{bb}$ of complex amplitude $b_{las}$ of this phonon laser is obtained from its correlation function as:

$$\begin{aligned} &\left\langle b_{las}(\tau)b_{las}^*(0)\right\rangle \approx \left|b_{las}\right|^2 e^{-i\omega_{las}\tau}\left\langle e^{-i(\phi'(t)-\phi'(0))}\right\rangle = \left|b_{las}\right|^2 e^{-\left\langle(\phi'(t)-\phi'(0))^2\right\rangle/2 - i\omega_{las}\tau} \\ &S_{bb}(\omega) = \int_{-\infty}^{+\infty} e^{-i\omega\tau}\left\langle b_{las}(\tau)b_{las}^*(0)\right\rangle d\tau = \left|b_{las}\right|^2 \frac{\left\langle\left(\phi'(1)-\phi'(0)\right)^2\right\rangle}{\left\langle\left(\phi'(1)-\phi'(0)\right)^2/2\right\rangle^2 + (\omega+\omega_{las})^2} \end{aligned} \tag{S29}$$

Therefore, the linewidth of the phonon laser $\Delta v = \frac{1}{2\pi}\left\langle\left(\phi'(1)-\phi'(0)\right)^2\right\rangle = \frac{\gamma_0(n_{th}+1/2)P_{eff}}{4\pi n_L}$ . The relation between Allan deviation $\sigma(\tau)$ and the phonon laser linewidth $\Delta v$ is $\sigma(\tau) = \sqrt{\frac{\Delta v}{2\pi\tau}}$ by considering Eq. (S23) and Eq. (S26). We note that the Allan deviation and linewidth representations are appropriate for characterizing white noise frequency jitter. For the non-white noise, the frequency-dependent $S_{ff}$ offers a more convenient representation.

In this study, the normalized SNR for nanomechanical frequency detection is defined by comparing the NEPs against a single-mode phonon laser, whose responsivity is unity and whose fundamental frequency-noise PSD is $S_{ff} = (2\pi)^{-2}\left(\gamma_0(n_{th}+1/2)/2n_L\right)$. Quantitatively, combined with Eq. (S25), the normalized SNR for multimode lasers is given by

$$SNR_{normalized} = \left(\frac{\partial \omega_{las}}{\partial \Omega_0}\right)^2 \left(\left(\frac{\partial \omega_{las}}{\partial \Omega_0}\right)^2 + \left(\frac{\partial \omega_{las}}{\partial (\gamma_0 / 2)}\right)^2\right)^{-1} \tag{S30}$$

Therefore, the normalized SNR does not exceed unity, indicating that the NEP-based sensing performance cannot surpass that of the single-mode laser, provided that the phonon number $n_L$ in resonator $b_0$ is the same. Here the normalized SNR is unity for the optimal protocol, for instance, if $\partial \omega_{las} / \partial \gamma_0$ is zero.

### B. Discussion of fundamental sensitivity near LEPs

Although we focus on the fundamental sensitivity near NEPs in the main text, we note that the equivalence between the fundamental noise and the parametric fluctuations shown in Eq. (S19) is valid in the linear regime for LEPs. For instance, the equivalent frequency fluctuations $\sqrt{\gamma_0 / 2}Y / \sqrt{n_L}$ are also mixed with the signal associated with mechanical-frequency variations. The contribution of equivalent dissipation fluctuations $\sqrt{2\gamma_0}X / \sqrt{n_L}$ can be filtered out in an optimal protocol by adjusting the quadrature angle of *I–Q* demodulation under coherent driving. Thus, for this case, the conclusion regarding the absence of fundamental sensitivity enhancement near NEPs is directly transferable to LEPs. Next, we systematically discuss several more general cases based on the equivalence between fundamental noise driving and the corresponding parametric jitter in the dynamical matrix.

*Nonstationary protocols*: For some protocols where $n_L$ is time-dependent (e.g., when the driving frequency is swept across the cavity), the frequency fluctuations are no longer stationary. Therefore, the PSD $S_{ff}$ is no longer suitable for characterizing the fundamental mechanical frequency noise. Here we assume the signal varies sufficiently slowly to be regarded as constant over $[t, t + \Delta t]$. We focus on the equivalent frequency fluctuations within this interval. The frequency fluctuations at different times *t* are weighted to obtain the minimum-variance unbiased estimator of the signal. Using the method of Lagrange multipliers, the optimal weights are found to scale with $n_L(t)$, which is directly measurable. Therefore, following Eq. (S19), the time-averaged frequency fluctuations over $[t, t + \Delta t]$ with optimal weighting follow a normal distribution $N\left(0, \gamma_0 n_{th} / \left(2\langle n_L(t)\rangle \Delta t\right)\right)$. $\langle n_L(t)\rangle$ is the interval-averaged phonon number over $[t, t + \Delta t]$, and it becomes nearly constant and independent of the selected interval for sufficiently large $\Delta t$. Thus, the minimum-variance unbiased estimate of the mechanical frequency depends on the average phonon number $\langle n_L(t)\rangle$ and is not directly related to the EP concept, its order, or a specific detection method. Consequently, under these circumstances, the root response near LEPs does not lead to an improvement in fundamental sensitivity.

*Generalized signal and noise loading*: In the above discussion of the fundamental sensitivity near EPs, the signal and noise are loaded onto the parameters of a single physical resonator. We now extend the discussion to systems involving multiple physical resonators, where multiple signal and noise channels may be present. In addition, the modes considered are not limited to phonon modes. Similar to Eq. (S18), the dynamics of each mode can be expressed as:

$$\frac{d\hat{b}_j}{dt} = -i\left(\Omega_j + A_\Omega^j \lambda(t) + \Delta\Omega_j\right)\hat{b}_j - \left(\gamma_j + A_\gamma^j \lambda(t) + \Delta\gamma_j\right)\hat{b}_j / 2 + \sum_{k\neq j} J_{jk}\hat{b}_k \tag{S31}$$

Here, $\Delta\Omega_j$ and $\Delta\gamma_j$ are equivalent parametric variations induced by fundamental noise. The terms $A_\Omega^j \lambda(t)$ and $A_\gamma^j \lambda(t)$ represent the system-parameter variations induced by the signal $\lambda(t)$. As a result, the weighted measurement output takes the form:

$$\sum_{j=0}^{n} w_\Omega^j \left(A_\Omega^j \lambda(t) + \Delta\Omega_j\right) + \sum_{j=0}^{n} w_\gamma^j \left(A_\gamma^j \lambda(t) + \Delta\gamma_j\right) \tag{S32}$$

Therefore, the optimal signal-to-noise ratio is obtained by optimizing the weight vector $\tilde{w} = (w_\Omega^0, \cdots, w_\Omega^n, w_\gamma^0, \cdots, w_\gamma^n)$ as:

$$\min_{\tilde{w}} \frac{\tilde{w} M \tilde{w}^T}{\left(\tilde{w}\tilde{A}^T\right)^2} \tag{S33}$$

Here, $\tilde{A} = \left(A_\Omega^0, \cdots, A_\Omega^n, A_\gamma^0, \cdots, A_\gamma^n\right)$, and $M$ is the symmetrized covariance matrix of the parametric jitter $M_{ij} = \left\langle \tilde{\xi}_i\tilde{\xi}_j + \tilde{\xi}_j\tilde{\xi}_i \right\rangle / 2$ with $\tilde{\xi} = \left(\Delta\Omega_0, \cdots, \Delta\Omega_n, \Delta\gamma_0, \cdots, \Delta\gamma_n\right)$. Here $\tilde{w}\tilde{A}^T$ is the signal response, whereas $\tilde{w}M\tilde{w}^T$ is the noise variance. It should be noted that only the relative weights of different signal and noise channels matter in this optimization problem, whereas the absolute scale is irrelevant. Consequently, the divergence of responsivity near EPs, which corresponds to large absolute values of some weight components, does not lead to a fundamental enhancement of the optimal SNR.

Beyond this fundamental consideration, there is also a practical obstacle: when multiple physical resonators are coupled, the responses to the different components of $\tilde{\xi}$ are mixed together.

In this situation, it is difficult to realize the optimal $\tilde{w}$. By contrast, when the physical resonators are decoupled from each other and measured independently, the relative weight of frequency jitter and linewidth jitter at each resonator can be freely tuned by balanced homodyne detection. In addition, the relative weight among different resonators can also be freely optimized through a multi-channel data fusion process to reach the fundamental sensitivity limit.

For some protocols, the signal is loaded through the coupling element as $\Delta J_{jk}$. In this case, $\Delta J_{jk}\hat{b}_k = \Delta J_{jk}\langle b_k\rangle / \langle b_j\rangle \hat{b}_j$, thus the off-diagonal element perturbation $\Delta J_{jk}$ is equivalent to a complex diagonal variation $\Delta J_{jk}\langle b_k\rangle / \langle b_j\rangle$, reducing to the case we discussed above. As a simple case, the coupling jitter between two modes can be regarded as the frequency jitter of collective modes after basis transformation.

*Vacuum-field interference in the readout:* According to input-output theory, the fluctuation field $\delta b_{out}^0 = \delta b_{in}^0 - \sqrt{\gamma_{ex}}\delta b_0$ ( $\hat{b}_0$ is not restricted to a phonon mode). This suggests that the cavity noise fluctuations alone do not fully describe the output noise, as the input vacuum fields $\delta b_{in}^0$ are also directly reflected. However, for an active system, the noise fluctuations $\delta b_0$ are dramatically amplified, especially near EPs. As a result, the reflected field $\delta b_{in}^0$ is negligible compared with $\sqrt{\gamma_{ex}}\delta b_0$ in the detected output field $\delta b_{out}^0$. This approximation also holds for other cases, such as thermally dominated fluctuations with vacuum-noise-limited detection in our system.

In contrast, for a passive system, this reflected vacuum field $\delta b_{in}^0$ interferes with the leakage field $\sqrt{\gamma_{ex}}\delta b_0$ from the cavity mode $\hat{b}_0$ to generate an output vacuum field $\delta b_{out}^0$. This indicates that the contribution of the directly reflected field $\delta b_{in}^0$ is non-negligible. Here $\delta b_0 = R \times \sqrt{\gamma_0}\delta b_{in}^0 / \langle b_0\rangle$, where $R$ is the complex responsivity shared by the equivalent complex frequency jitter $\delta b_{in}^0 / \langle b_0\rangle$ and the co-amplified signal at the detection port. As the output fluctuation remains a vacuum field, it imposes the constraint $\left|1 - R \times \sqrt{\gamma_0} / \langle b_0\rangle\right| = 1$ for a single cavity mode with $\gamma_0 = \gamma_{ex}$. This means that the maximum amplitude of $R$, and hence the maximum SNR, is reached when $R \times \sqrt{\gamma_0} / \langle b_0\rangle = 2$. Notably, the leakage field is suppressed to half of its original amplitude by the reflected field. This condition is satisfied when the coherent drive is resonant with the cavity mode $\hat{b}_0$. Moreover, the corresponding SNR reaches the Cramér–Rao bound[17]. When multiple modes interact to form the EPs, noise from other channels also enters the detection port, contributing to the output vacuum field. As a result, $\left|1 - R \times \sqrt{\gamma_0} / \langle b_0 \rangle\right| < 1$ and $\left|R \times \sqrt{\gamma_0} / \langle b_0 \rangle\right| < 2$. Therefore, the introduction of EPs does not lead to fundamental sensitivity enhancement.

## VI. ADDITIONAL EXPERIMENTAL RESULTS FOR PHONON LASER LINEWIDTH BROADENING

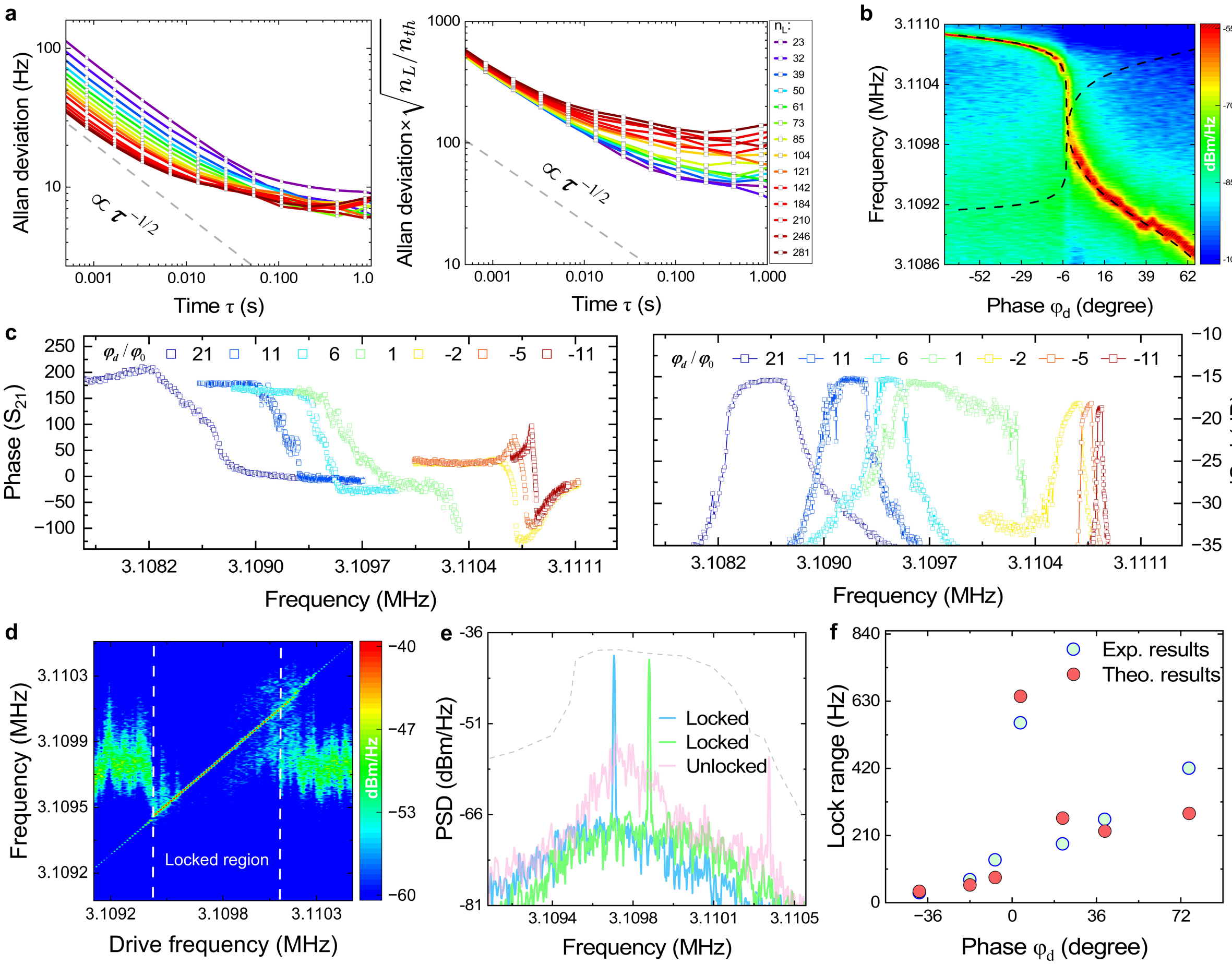

**Fig. S8. Additional experimental results for linewidth broadening analysis of the phonon laser.** (**a**) Frequency Allan deviation for a single-mode phonon laser measured at multiple phonon amplitudes, together with its normalized form after removing phonon-energy dependence. Gray dashed lines represent a reference slope of -1/2 in the log–log plot. (**b**) Colormap of the BHD signals measured at multiple delay phases $\varphi_d$ near $NEP_3$, with dashed lines representing the corresponding theoretical eigenvalues. (**c**) Phase and amplitude of $S_{21}$ response measured during the injection-locking experiment at multiple $\varphi_d$ values near $NEP_3$. (**d**) Colormap of the BHD signals for multiple drive frequencies in the injection-locking experiment at $\varphi_d$ = 3.5849° near $NEP_3$. (**e**) Corresponding PSDs in the locked and unlocked regions of (d). (**f**) Comparison between the theoretical and experimental locking ranges near $NEP_3$. In (c), the delay phases are represented in units of $\varphi_0 = 3.5849°$, corresponding to the additional phase induced by one FPGA clock period.

Since the fundamental linewidth broadening of the phonon laser arises from the thermal force or quantum fluctuation-induced phase noise, we characterized the Allan deviation experimentally using the simplest single-mode phonon laser to reveal the influence of fundamental phase noise. As shown in Fig. S8a, the slope of frequency Allan deviation $\sigma$ versus time $\tau$ in the log-log plot is close to -1/2 for various phonon numbers $n_L$. Moreover, the normalized Allan deviations $\sigma\sqrt{n_L / n_{th}}$ converge to the same value at short averaging times. These results are consistent with the theoretical model in Section V and indicate that phase noise is the dominant noise source at short times $\tau$.

As discussed in Section V, the thermally driven term is effectively amplified in the phase equation near NEPs, leading to linewidth broadening. Therefore, if the phonon laser is driven by a single-frequency signal rather than by thermal noise, the effective driving strength of the input signal will also be enhanced. This analogy indicates that the phase of the phonon laser can follow the drive signal, enabling injection-locking at very low drive power for a fixed detuning $\Delta_\omega$ between the driving and phonon laser frequencies. Thus, the same mechanism responsible for the free-running linewidth broadening also broadens the injection-locking range, as described by the phase equations:

$$\frac{d\tilde{\phi}}{dt} - \Delta_\omega = -\frac{\sqrt{\gamma_0}\left|b_{drive}\right|}{\left|b_{las}\right|}\mathrm{Im}\left(\frac{\left\langle \Psi_{las}^L \middle| e_0 \right\rangle}{\left\langle \Psi_{las}^L \middle| \Psi_{las}^R \right\rangle} e^{i\tilde{\phi}}\right) \tag{S34}$$

where $\tilde{\phi}$ is the relative phase of the phonon laser with respect to the phase of the driving force, and $\Delta_\omega = \omega_{las} - \omega_{drive}$. Here, we assume the driving force acts on the real resonator. The relative phase $\tilde{\phi}$ will be locked provided that $d\tilde{\phi} / dt = 0$ and the stability condition $\partial(d\tilde{\phi} / dt) / \partial\tilde{\phi} < 0$ is satisfied. Thus, phase locking occurs for detuning satisfying $\Delta_\omega \le \sqrt{\gamma_0}\left|\frac{\left\langle \Psi_{las}^L \middle| e_0 \right\rangle b_{drive}}{\left\langle \Psi_{las}^L \middle| \Psi_{las}^R \right\rangle b_{las}}\right|$, and the divergence of the term $\left\langle \Psi_{las}^L \middle| \Psi_{las}^R \right\rangle^{-1}$ near the LEP leads to an enlarged locking range around the NEP, as the NEP is close to the LEP.

Next, the injection-locking behavior near $NEP_3$ is experimentally observed. As shown in Fig. S8b, the delay phase is tuned close to the $NEP_3$. Subsequently, a single-frequency drive is added, along with the feedback, to modulate the EOM. $|S_{21}|$ is dramatically amplified in the locking regime, where the phonon laser follows the external drive. Accordingly, the measured $S_{21}$ response (Fig. S8c) shows a broad locking range near $NEP_3$. In addition, the phase of $S_{21}$ shifts by about 180 degrees across the locking region, in agreement with the conventional theoretical description. The detailed injection-locking spectra are shown in Fig. S8d–e. Despite the linewidth broadening of the free-running phonon laser, it exhibits an ultra-narrow linewidth in the injection-locking range, as its phase follows that of the single-frequency drive. Furthermore, we define the locking range as the 3-dB bandwidth of $|S_{21}|$, and the experimental and theoretical results shown in Fig. S8f are consistent. We note that the $|S_{21}|$ curve near $NEP_3$ is not entirely flat in the locked region, which is distinct from single-mode injection locking. The slight discrepancy between the experimental results and the theoretical prediction is likely due to the omission of other non-lasing modes in the multimode system.

## VII. ILLUSION OF SENSITIVITY ENHANCEMENT NEAR NEPs

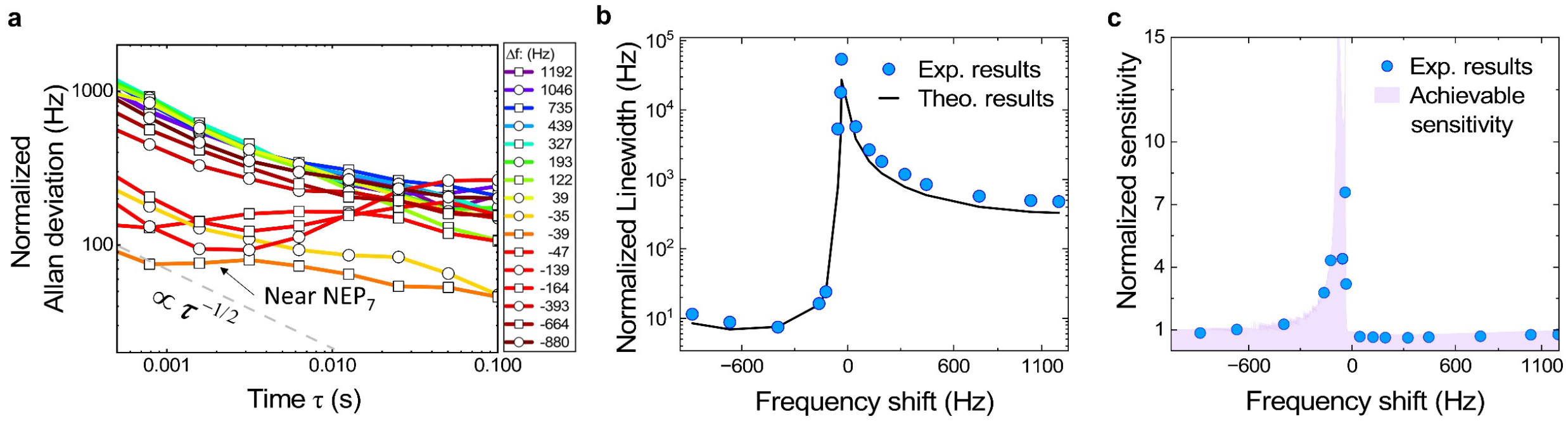

**Fig. S9. The illusion of sensitivity enhancement near NEPs.** (**a**) Normalized Allan deviation for the data in Fig. 4a. (**b**) and (**c**) compare the corresponding experimental and theoretical results for the normalized linewidth and sensitivity, respectively, under the assumption of white-noise frequency fluctuations.

Next, we assume that the frequency jitter behaves as white frequency noise to demonstrate the illusion of sensitivity enhancement near $NEP_7$. The normalized Allan deviation $\sigma(\tau)\sqrt{n_L / n_{th}} / (\partial\omega_{las} / \partial\varepsilon)$, shown in Fig. S9a, is obtained by comparing the Allan deviation $\sigma(\tau)$ and the responsivity $\partial\omega_{las} / \partial\varepsilon$ at the same average phonon number $n_L$ of the nanomechanical resonator. For short averaging time $\tau$, the normalized Allan deviation near the $NEP_7$ (orange line) is lower, seemingly indicating enhanced sensitivity. Meanwhile, the phase-noise-induced normalized linewidth $\Delta\nu \times n_L / n_{th}$ is shown in Fig. S9b. It is deduced from the Allan deviation at short averaging time $\tau = 0.786$ ms, during which a -1/2 slope is maintained, showing good agreement with the theoretical linewidth. Furthermore, the measurement sensitivity near NEP$_7$, normalized to the fundamental sensitivity of a single-mode laser as $\sqrt{\frac{\gamma_0 n_{th}}{2 n_L \tau} \frac{\partial\omega_{las} / \partial\varepsilon}{2\pi\sigma(\tau)}}$, is shown in Fig. S9c $(\tau = 0.786\ ms)$. It appears enhanced near NEPs, consistent with the theoretical trend.

Nevertheless, the Allan deviation will deviate from the -1/2 slope in the log-log plot once the nonlinear feedback responds to thermal noise, leading to colored frequency fluctuations. Therefore, the noise behavior differs between short and long averaging times $\tau$, since $\sigma(\tau)\sqrt{\tau}$ is no longer $\tau$-independent. As a result, combining the static responsivity $\partial\omega_{las} / \partial\varepsilon$ and the short-$\tau$ deviation $\sigma(\tau)$ to derive the sensitivity is inappropriate. We emphasize that the feedback-strength response induces non-white (colored) noise fluctuations. Ignoring this contribution would result in the illusion of sensitivity enhancement near NEPs.

## VIII. VERIFICATION OF SENSITIVITY UNDER FUNDAMENTAL NOISE

### A. Sensitivity verification via Simulink simulations

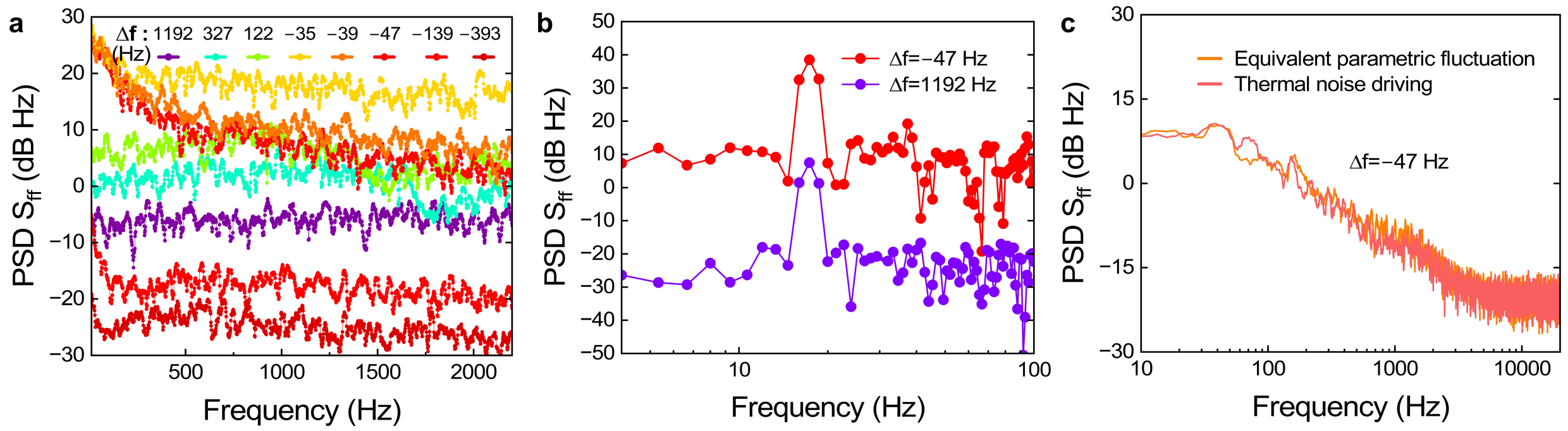

**Fig. S10. Simulated results of fundamental noise induced frequency fluctuations.** (**a**) Simulated PSD $S_{ff}$ of the instantaneous mechanical lasing frequencies, extracted from the projection $\langle \Psi_L^{las} | \Psi \rangle$. The $\Delta f$ values are taken from Fig. 4b. (**b**) Simulated PSD $S_{ff}$ after applying a 10-Hz perturbation $\varepsilon$ to the frequency of $b_0$, using a subset of the parameters in (a). The ratio $n_{th} / n_L$ is held constant for all $\Delta f$ values. (**c**) Comparison of the simulated PSD $S_{ff}$ between the thermal-noise-driving model and the equivalent parametric-fluctuation model.

To further validate the sensitivity analysis in the main text, we use Simulink to simulate the frequency jitter near NEPs and compare it with the experimental and theoretical results. The simulated PSD $S_{ff}$, obtained using the same parameters as those in Fig. 4b, is presented in Fig. S10a. The simulated results in Fig. S10a agree well with the experimental results in Fig. 4b. Moreover, frequency-dependent features in $S_{ff}$ become prominent near NEPs. In the low-frequency regime, the noise is significantly higher than that at high frequencies, where the phase random walk dominates. This suggests that the traditional Petermann factor, which quantifies the phase noise level, becomes unreliable for sensitivity analysis near NEPs. Accordingly, the perturbation $\varepsilon$ is applied to the frequency of mode $b_0$, with the temperature fixed at 2.93 K to ensure the perturbation signal remains well above the noise floor in a short-time simulation window. The simulated PSD in Fig. S10b shows that the increase in noise suppresses the expected SNR enhancement near NEPs, with no indication of sensitivity enhancement near NEPs. In addition, we theoretically demonstrate in Section V that thermal noise driving is equivalent to independent fluctuations in the mechanical frequency and dissipation rate. The comparison between the simulation results of the thermal noise driving model and the equivalent parameter fluctuation model is further illustrated in Fig. S10c, and their PSDs $S_{ff}$ agree well.

### B. Generation of periodic perturbation

In the experiment investigating the fundamental sensitivity near NEPs, we replaced the nanomechanical resonator with the computational synthetic resonator. To control the frequency of the synthetic resonator, an external signal with a periodically modulated frequency was applied to tune the FPGA clock. The clock frequency variation rescales the responses of the synthetic resonator in the frequency domain, thereby resulting in a frequency shift of $b_0$. As shown in Fig. S11, the central frequency of the PSD associated with $b_0$ shifts linearly with the frequency of the external clock signal (near 10 MHz).

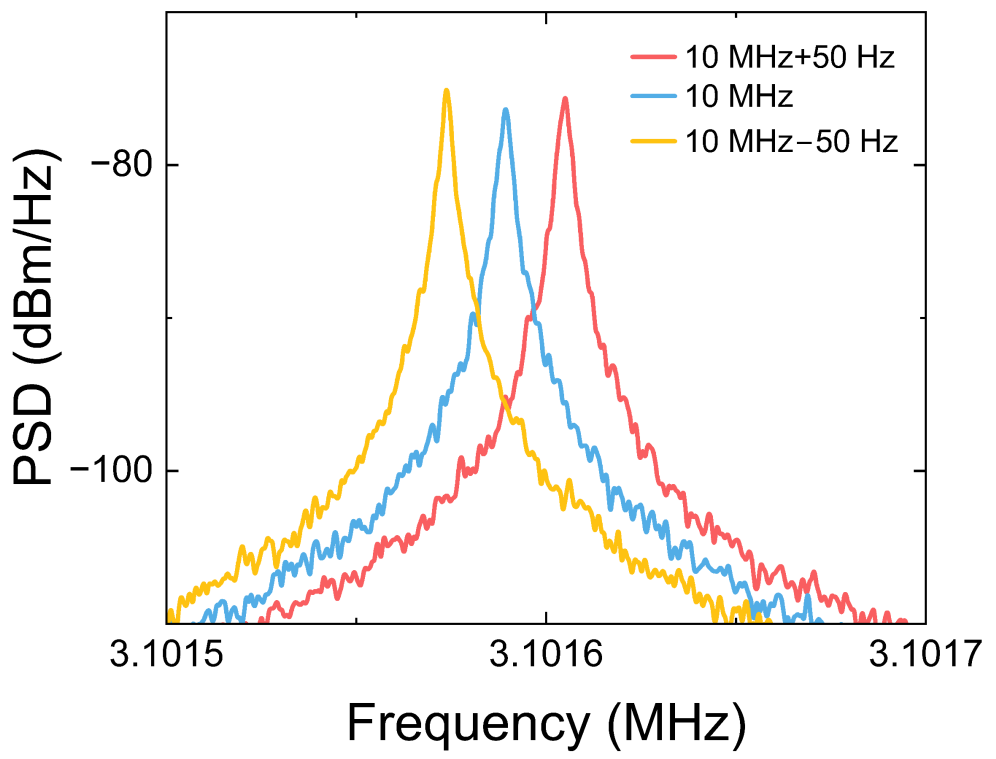

**Fig. S11.** Displacement PSD of the mode $b_0$ under different clock frequencies with the nanomechanical resonator replaced by a computational synthetic resonator.

## C. Fundamental sensitivity in Hatano–Nelson–type model

Typically, the measurement sensitivity near NEPs is overestimated if the dynamics of the long-term nonlinear feedback is neglected. As an exception, the Petermann factor alone in Supplementary Ref. 14 correctly explains the absence of sensitivity enhancement near $EP_2$[18], because the nonlinearity of the Brillouin gain does not alter the lasing frequency in that system. Here, we demonstrate an analogous behavior in the Hatano–Nelson–type model. The effective Hamiltonian of the unidirectional Hatano–Nelson–type model can be expressed as:

$$H_{eff} = (\Omega_0 - i\gamma_0 / 2) I + J_H e^{i\varphi_d / n} \hat{S} \tag{S35}$$

where $n$ is the number of resonators and $\hat{S}$ is the cyclic permutation operator. The eigenvalues of this unidirectional phonon laser are given by $\omega = \Omega_0 - i\gamma_0 / 2 + J_H e^{i(\varphi_d / n + 2\pi m / n)}$. The coupling strength $J_H e^{i(\varphi_d / n + 2\pi m / n)}$ is tuned to be purely imaginary by adjusting the delay phase $\varphi_d$ and approaches $i\gamma_0 / 2$ at equilibrium. When the complex fluctuation $\sqrt{\gamma_0 / 2}\dfrac{-Y + iX}{\langle |b_0| \rangle}$ is introduced into one of the resonators, the lasing-mode frequency is $\omega' \approx \omega + \sqrt{\gamma_0 / 2}\dfrac{-Y + iX}{n\langle |b_0| \rangle}$. Then the coupling strength $J_H e^{i\varphi_d / n}$ varies to compensate for the fluctuation in the imaginary part of $\omega'$. As a result, the feedback has a negligible influence on the lasing frequency $\mathrm{Re}(\omega')$ over long timescales. This implies that the fundamental frequency-noise floor at low frequencies is the same as that at high frequencies. Theoretically, the fundamental noise contribution to the phonon-laser linewidth $\Delta\nu$ is $\frac{\gamma_0 n_{th}}{4\pi n_L}\frac{1}{n^2}$, which is inversely proportional to the square of the resonator number $n$. This scaling law is shown experimentally in Fig. S12a. Moreover, the responsivity $\partial\omega_{las}/\partial\varepsilon$ of the lasing mode scales with $1/n$. This behavior is reflected in the low-frequency region of the PSD $S_{ff}$ in Fig. S12b, which scales with $1/n^2$. Here, the technical resonant-frequency drift is the dominant contribution. Consequently, the fundamental sensitivity for such a system remains the same regardless of the resonator number $n$.

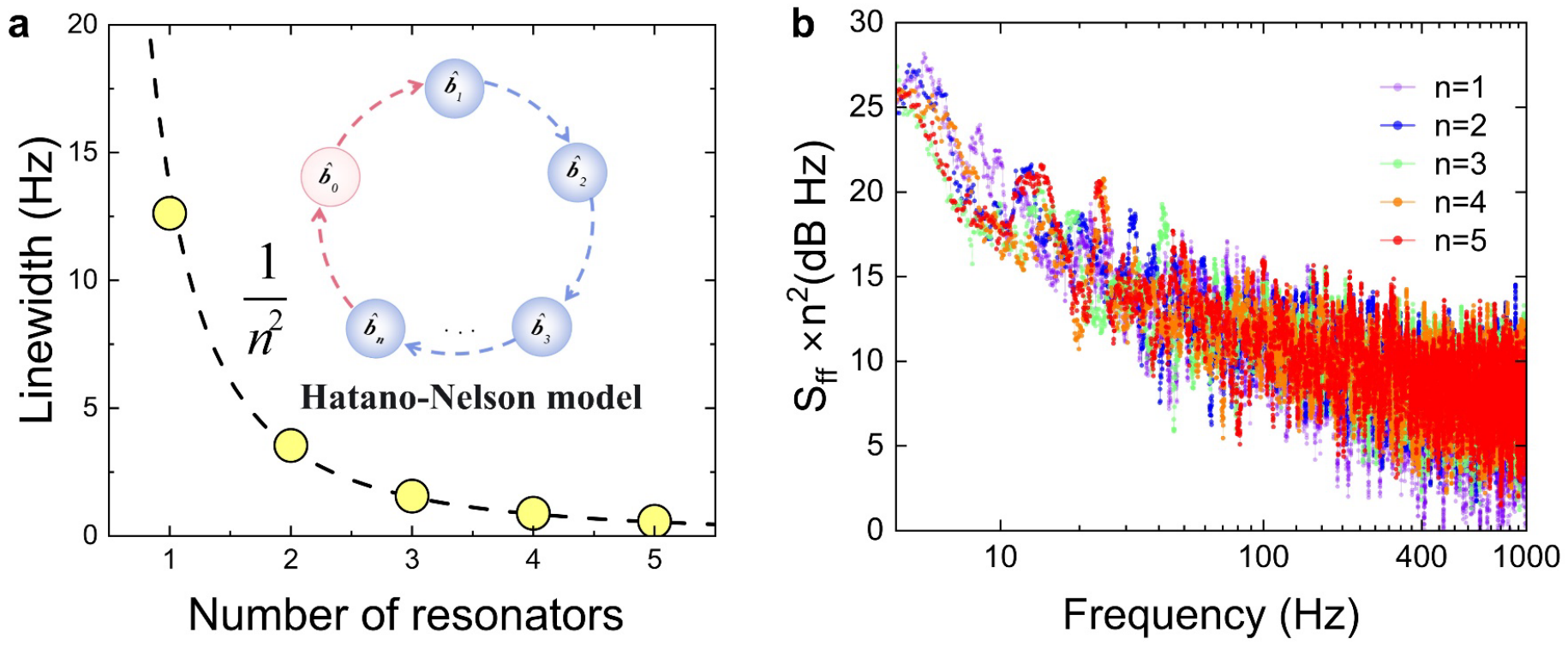

**Fig. S12. Experimental sensitivity analysis of the Hatano–Nelson–type model.** Linewidth (**a**) and corresponding rescaled PSD $S_{ff}\times n^2$ (**b**) for the unidirectional phonon laser with different numbers of resonators $n$. A nanomechanical resonator $b_0$ is used, and the instantaneous lasing frequency is extracted from the projection $\langle\Psi_L^{las}|\Psi\rangle$. The average phonon number $n_L$ is kept constant for all datasets.

## IX. LONG-TERM FREQUENCY DRIFT ANALYSIS

Noise in the computational system can be divided into two categories based on its consequences. The first category is fundamental noise (thermal and quantum noise). The influence of the fundamental noise has been well discussed in Sections V–VIII. The second type is technical noise. Here we focus on technical noise that induces drift in intrinsic system parameters, and leads to long-term lasing-frequency drift. In this system, the crucial parameters are the phase $\varphi_d$, which is determined by system delay; the feedback strength, which is influenced by BHD interference phase $\theta$, light intensity, and PD response; the intrinsic mechanical properties like the mechanical resonant frequency $\Omega_0$ and the mechanical linewidth $\gamma_0$; and the parameters of digital filters that represent the synthetic resonators' properties. In our analysis, the drift of the intrinsic mechanical frequency $\Omega_0$ is dominant. As this noise limits the precision for obtaining the responsivity in Fig. 3e of the main text, we replaced the real resonator with the synthetic one, and the corresponding experimental setup is shown in Fig. S13a. The delay controller (shift register) and the optical delay line are used to tune the system delay coarsely and finely, respectively. Evidence supporting the conclusion that intrinsic frequency drift is the dominant source is summarized below:

*Feedback delay-phase drift*: The error signal in the open-loop mode of the PID controller shows that the interference phase takes several seconds to drift by 2π. This indicates that the delay imprecision is on the order of $\lambda/c \sim 10\,fs$, which corresponds to a negligible drift of the delay phase $\varphi_d$. Moreover, for single-mode lasing, the lasing frequency is $\Omega_0 + (\gamma_0/2)\tan\varphi$ with $\varphi = \varphi_d - \pi/2$, ensuring the lasing mode has zero imaginary part. Therefore, the frequency drift scales with $d\tan\varphi/d\varphi = \sec^2\varphi$. For the real resonator $b_0$, the Allan deviation in Fig. S13b is independent of the phase $\varphi$ for the fixed phonon number $n_L$, indicating the drift of the delay phase $\varphi$ is negligible. In contrast, when $b_0$ is replaced by a computational synthetic resonator, the phase $\varphi$-dependent Allan deviation in Fig. S13b suggests the delay-phase drift may play a role once other contributions are suppressed.

*Feedback strength drift:* The feedback strength drift can result in variations in the gain of the mechanical lasing mode. Consequently, the mechanical amplitude will increase or decrease in response to changes in the mechanical gain. Over long timescales, the mechanical amplitude change compensates for the influence of the feedback drift. Thus, the frequency jitter resulting from the feedback strength drift will be negligible for all $\varphi$ settings over long averaging times $\tau$. Accordingly, the feedback strength drift cannot account for the non-negligible long-term Allan deviation in Fig. S13b.

*Intrinsic mechanical properties drift:* We monitored the long-term drift of both the intrinsic mechanical frequency and mechanical linewidth, as shown in Fig. S13c. Here, the maximum variations of intrinsic mechanical frequency and mechanical linewidth are 10 kHz and 136 Hz, respectively. Therefore, the long-term drift of the intrinsic frequency, as widely reported[19–21], is substantially larger than that of the linewidth. If the mechanical linewidth drift is dominant, the drift of the lasing frequency will scale with $\tan\varphi$ over long timescales. This contradicts the observation in Fig. S13b, where the Allan deviation is insensitive to $\varphi$ when the phase noise from thermal driving is no longer dominant.

*Parameter drift of synthetic resonators:* As the FPGA clock is highly stable, the parameter drift of synthetic resonators is negligible. This conclusion is supported by the comparison of Allan deviation with and without the use of the nanomechanical resonator in Fig. S13b. If needed, the system stability can be further improved by referencing an external atomic clock.

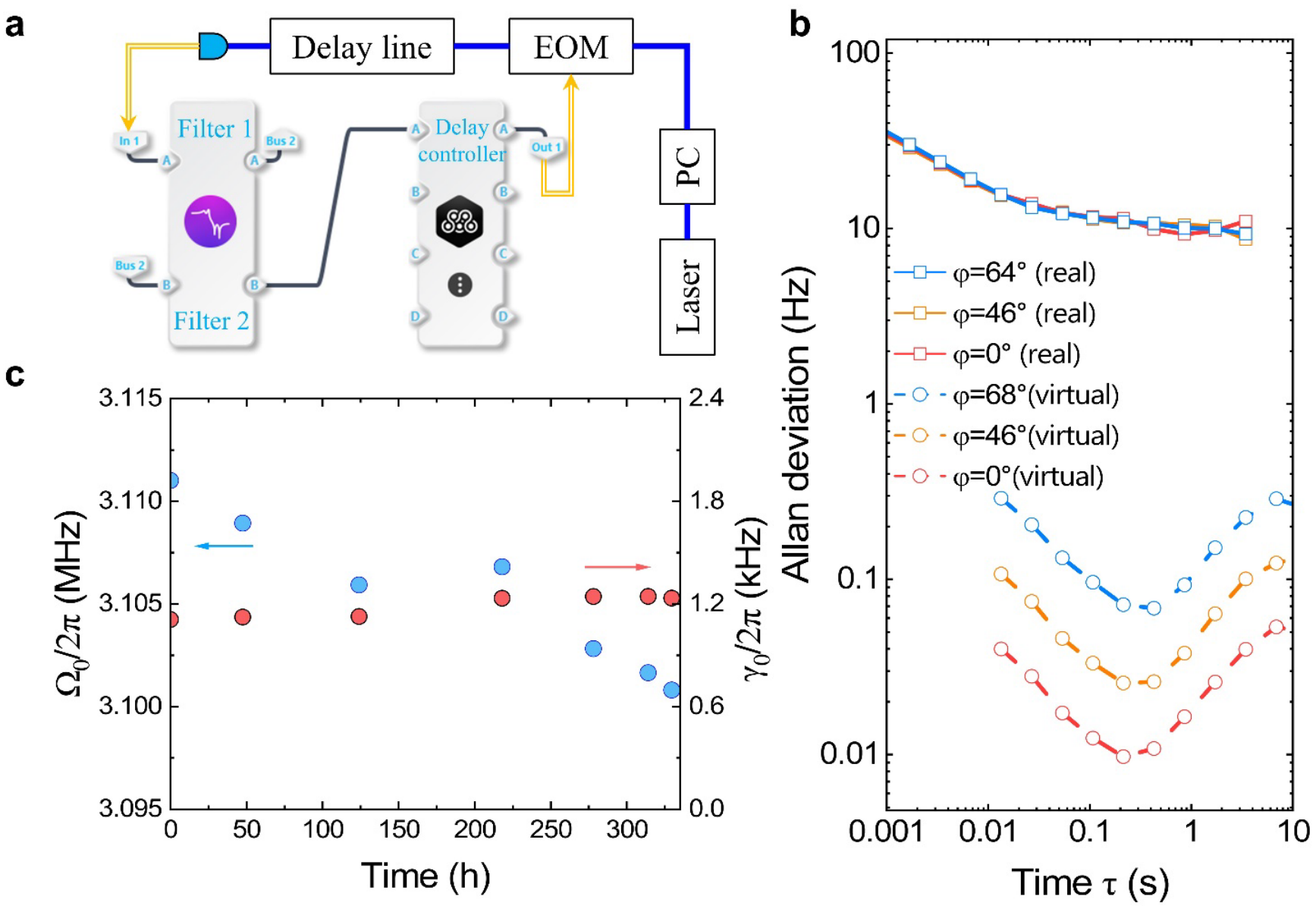

**Fig. S13. Long-term noise analysis.** (**a**) The setup for realizing NEPs without using the nanomechanical resonator. (**b**) The Allan deviation of $b_0$ at different phases $\varphi = \varphi_d - \pi/2$ in single-mode lasing. The dashed (solid) lines represent the experimental results obtained using the synthetic (nanomechanical) resonators. (**c**) The long-term drift of the intrinsic mechanical resonant frequency $\Omega_0$ and the mechanical linewidth $\gamma_0$.

## X. PARAMETER SETTINGS FOR REALIZING LEPs

LEPs in parity-time (PT) and anti-parity-time (anti-PT)-symmetric systems are demonstrated in Fig. 3a of the main text. Their effective Hamiltonians are given by $\begin{pmatrix} \Omega_0 + i\gamma_{gain}/2 & J \\ J & \Omega_0 - i\gamma_1/2 \end{pmatrix}_{PT}^{EP_2}$,

$\begin{pmatrix} \Omega_0 - i\gamma_1/2 & J & 0 \\ J & \Omega_0 - i\gamma_0/2 & J \\ 0 & J & \Omega_0 - i\gamma_2/2 \end{pmatrix}_{PT}^{EP_3}$, and $\begin{pmatrix} \Omega_0 - i\gamma_0/2 & iJ \\ iJ & \Omega_1 - i\gamma_1/2 \end{pmatrix}_{anti-PT}^{EP_2}$.

For the PT LEP$_2$, the linewidth of the synthetic resonator $\gamma_1$ is set to $\gamma_0$. The additional gain is applied to the real resonator, with $\gamma_{gain} = 0.41\gamma_0$. For the PT LEP$_3$, the dissipation rates of two other synthetic resonators are $\gamma_1 / 2\pi = 100$ Hz, and $\gamma_2 = 2\gamma_0 - \gamma_1$. For the anti-PT LEP$_2$, the frequency of the synthetic resonator $\Omega_1$ is set to $\Omega_0 - 5\gamma_0$.